\documentclass[preprint,12pt]{elsarticle}

\usepackage[margin=3.0cm]{geometry}

\usepackage{graphicx,subfigure}
\usepackage{amssymb}
\usepackage{amsmath}
\usepackage{bm}
\usepackage{lineno}
\usepackage{natbib}
\usepackage[squaren]{SIunits}
\usepackage{color}
\usepackage{multirow}
\usepackage[normalem]{ulem}
\usepackage{cancel}
\journal{ArXiv.org}

\definecolor{codegreen}{rgb}{0,0.6,0}
\definecolor{codegray}{rgb}{0.5,0.5,0.5}
\definecolor{codepurple}{rgb}{0.58,0,0.82}
\definecolor{backcolour}{rgb}{0.95,0.95,0.92}

\newcommand{\PP}[2]{\frac{\partial#1}{\partial#2}}

\begin{document}

\begin{frontmatter}

\title{A new field solver for modeling of relativistic particle-laser interactions using the particle-in-cell algorithm}

\author[UCLAEE]{Fei Li\corref{cor1}}
\author[UCLAPH]{Kyle G. Miller}
\author[SLAC]{Xinlu Xu\corref{cor1}}
\cortext[cor1]{Corresponding authors: lifei11@ucla.edu (Fei Li), xuxinlu@slac.stanford.edu (Xinlu Xu)}
\author[UCLAPH]{Frank S. Tsung}
\author[UCLAPH]{Viktor K. Decyk}
\author[BNU]{Weiming An}
\author[IST,ISCTE]{Ricardo A. Fonseca}
\author[UCLAEE,UCLAPH]{Warren B. Mori}

\address[UCLAEE]{Department of Electrical Engineering, University of California Los Angeles, Los Angeles, CA 90095, USA}
\address[UCLAPH]{Department of Physics and Astronomy, University of California Los Angeles, Los Angeles, CA 90095, USA}
\address[SLAC]{SLAC National Accelerator Laboratory, Menlo Park, CA 94025}
\address[IST]{GOLP/Instituto de Plasma e Fus\~ao Nuclear, Instituto Superior T\'ecnico, Universidade de Lisboa, Lisbon, Portugal}
\address[ISCTE]{ISCTE - Instituto Universit\'ario de Lisboa, 1649--026, Lisbon, Portugal}
\address[BNU]{Department of Astronomy, Beijing Normal University, Beijing 100875, China}

\begin{abstract}

A customized finite-difference field solver for the particle-in-cell (PIC) algorithm that provides higher fidelity for wave-particle interactions in intense electromagnetic waves is presented.  In many problems of interest, particles with relativistic energies interact with intense electromagnetic fields that have phase velocities near the speed of light. Numerical errors can arise due to (1)~dispersion errors in the phase velocity of the wave, (2)~the staggering in time between the electric and magnetic fields and between particle velocity and position and (3)~errors in the time derivative in the momentum advance.  Errors of the first two kinds are analyzed in detail. It is shown that by using field solvers with different $\bm{k}$-space operators in Faraday's and Ampere's law, the dispersion errors and magnetic field time-staggering errors in the particle pusher can be simultaneously removed for electromagnetic waves moving primarily in a specific direction. The new algorithm was implemented into \textsc{Osiris} by using customized higher-order finite-difference operators. Schemes using the proposed solver in combination with different particle pushers are compared through PIC simulation. It is shown that the use of the new algorithm, together with an analytic particle pusher (assuming constant fields over a time step), can lead to accurate modeling of the motion of a single electron in an intense laser field with normalized vector potentials, $eA/mc^2$, exceeding $10^4$ for typical cell sizes and time steps.

\end{abstract}


\end{frontmatter}


\section{Introduction}
\label{sect:intro}

The interaction of relativistic charged particles with laser fields has attracted extensive attention in plasma and accelerator physics.  Examples of current research in frontier areas in which relativistic wave particle interactions are important include plasma-based acceleration of electrons/positrons and ions~\cite{tajima1979,chen1985,joshi2003,esarey2009,macchi2013,lu2007}, direct laser acceleration~\cite{pukhov1999}, quantum electrodynamic laser-plasma interactions~\cite{vranic2016}, free-electron lasers~\cite{pellegrini2016} and stochastic wave-particle interactions~\cite{mendonca1983,forslund1985,sheng2002,may2011,kemp2014}. The particle-in-cell (PIC) algorithm~\cite{dawson1983,hockney1988,birdsall2005} has been used for nearly half a century to study how plasmas and beams interact with radiation. It has also become a  powerful tool for modeling a variety of plasma and beam physics  processes.  Most current electromagnetic PIC codes use the finite-difference time-domain (FDTD) method as it is simple, versatile and straightforward to parallelize. The grid-based FDTD method discretizes the time-dependent Maxwell's equations using a central-difference approximation for both space and time domains. The resulting discretized set of equations is solved in a leapfrog manner in time, with the electric and magnetic field components interlaced in space when using the Yee mesh grid~\cite{yee1966}. Many numerical issues can arise due to the discretization, requiring careful use to avoid subtle spurious effects. Examples of known issues include improper numerical dispersion, numerical Cerenkov radiation and the associated numerical Cerenkov instability (NCI)~\cite{godfrey2013,xu2013,yu2015a,yu2015b}, finite-grid instability~\cite{langdon1970,okuda1972,meyers2015,huang2016} and numerical errors in the fields that surround relativistic particles~\cite{xu2019}. These errors do not always decrease proportionately with decreasing cell size and time step, making it important to deeply understand the cause of these effects in order to most efficiently remedy them.

Generally, the impact of numerical issues is problem-specific, and in many cases, no single algorithm can solve all problems. In this article, we consider intense laser fields interacting with particles that co-propagate with the laser fields at speeds close to the speed of light.  This situation arises in high-intensity laser-plasma interactions and plasma-based acceleration. It has been recognized for some time that errors arise when computing the trajectory of single particles in the fields of intense light waves (lasers)~\cite{lehe2014,arefiev2015,gordon2017}. In this work, we analyze several reasons for these errors and propose a solution that can be implemented into PIC codes that utilize finite-difference and FFT-based algorithms~\cite{dawson1983, birdsall2005}.

As we will show, the dominant error is often due to numerical dispersion. The time-space discretization causes an electromagnetic wave to propagate across the grid with errors in its dispersion relation that vary between Maxwell solvers and can depend on time step and cell size. These errors are of large concern particularly when the particle is co-moving with the laser close to the speed of light. In such a scenario, small errors in the phase velocity can lead to large differences in the resonant interactions between waves and particles. The trajectory of particles in phasespace is therefore very sensitive to these numerical errors. The spectral method~\cite{dawson1983, godfrey2014,yu2014,yu2015a,lehe2016}, \emph{i.e.}, solving Maxwell's equations in Fourier space, can remove  numerical errors due to spatial derivatives. Some refer to this as a pseudo-spectral method when grids are used. Furthermore, one can use these methods to exactly integrate the fields forward in time assuming the current is constant during a time interval (time step). This method, called the pseudo-spectral analytical time-domain (PSATD) method \cite{haber1973,  birdsall2005}, can thus provide a numerical-dispersion-free scheme for light propagating in vacuum. However, the PSATD method is not free from spurious effects when particles are included. Another advantage with FFT-based methods is that the entire algorithm improves in accuracy as the time step is reduced, including the particle advance. Therefore, convergence can be investigated by reducing the time step while keeping cell size fixed. While the use of FFT-based solvers can improve dispersion, they do so at a cost of decreased computational efficiency and parallel scalability unless a local FFT-based approach is used \cite{vay2013}. However, many existing codes are based on finite-difference methods, and shifting these to FFT-based  algorithms can require major changes to the software. Therefore, an FDTD method is desired that exhibits good dispersion characteristics and that improves in accuracy when the time step is reduced while keeping the cell size fixed.

The second important numerical issue specific to relativistic particle-laser interaction is the inaccurate evaluation of the Lorentz force during the particle advance. This inaccuracy is caused by the time staggering (by a half time step) between the electric and magnetic field components for time centering of the field equations.  In reality, the electric and magnetic fields in a plane wave are exactly in phase (and equal in amplitude when in vacuum for cgs or normalized units), so an ultra-relativistic particle in a co-propagating laser feels nearly vanishing transverse Lorentz force: the force is proportional to $(1-\frac{v_z}{c})\bm E_{\perp} \approx \frac{1}{2\gamma^2}\bm E_{\perp}$, where $\bm E_\perp$ is the transverse electric field, $v_z$ the longitudinal velocity of the particle, $\gamma$ the Lorentz factor and $c$ the speed of light. However, due to the time staggering of electromagnetic fields, at the time step when the electric force is known, the magnetic force must be approximated from the adjacent half time steps where the magnetic fields are defined. The time staggering leads to numerical errors larger than $1/2\gamma^2$ so that they dominate the Lorentz force felt by the particle. To solve this problem, a higher-order interpolation in time has been proposed~\cite{lehe2014} to approximate the magnetic force. However, this method has limitations for improving the accuracy of the Lorentz force evaluation and requires extra memory to store the fields for interpolation. Alternatively, the PSATD method could be formulated without time staggering to make it free of both numerical dispersion and errors in the Lorentz force.

In this article, we present a finite-difference (FD) based algorithm that simultaneously eliminates numerical dispersion along one direction and corrects for errors in the $\bm v \times \bm B$ force from the time staggering of the fields. This is done by first identifying the desired $\bm k$-space operators for the curl operations in Ampere's and Faraday's laws (these operators can easily be used in an FFT-based solver and adapted for any time step), then using the method described in Ref.~\cite{li2017} to construct a customized FD solver that replicates the desired $\bm k$-space operators. The solver also includes a correction to the current in order to guarantee that Gauss's law is satisfied at each time step.

A third numerical issue is the inaccuracy of the particle pusher. The Boris pusher~\cite{boris1972,qin2013} uses a second-order (leapfrog) operator for the time derivative and a split operator for the electric force and rotation from the $\bm v \times \bm B$ force. The momentum is advanced a half time step from $\bm E$, then rotated a full-time step from  $\bm v \times \bm B$, and then advanced a second half step from $\bm E$. The rotation can be done exactly with only small adjustments~\cite{boris1972,birdsall2005}. The main source of the error in the Boris push is that $\bm v$ is not known at the correct half time step, so an average is used. In addition, a relativistic code has $\bm v = \bm p/ m \gamma$, so there are several choices for defining $\bm v$ during the rotation since neither $\bm p$ nor $\gamma$ are known at the half step. Recently, there have been several ideas for improving on the Boris pusher; some of these were motivated to model the motion of charged particles in high-amplitude laser fields. Vay~\cite{vay2008} and Higuera and Cary (HC)~\cite{higuera2017} suggested using different definitions for $\gamma$ during the magnetic field rotation. Recently, Arefiev et al.~\cite{arefiev2015} proposed using a sub-cycling technique when the fields were very large, while Gordon et al.~\cite{gordon2017} showed that a covariant pusher could be exact if the fields are constant during a proper time step. Very recently, P\'etri~\cite{petri2019} proposed an exact or analytic pusher (for constant fields over a time step) in which a mapping between the proper and lab time for each particle is required. These analytic pushers do not use analytic results for the position advance. The sub-cycling method essentially recovers the analytic result when small enough time steps are used. We have implemented the HC and an extension of the ideas of the Gordon and P\'etri pushers into \textsc{Osiris}~\cite{fonseca2002}. We find that when combined with our proposed solver, the HC pusher agrees very well with the analytic pusher (and theory) until the laser strength parameter $a_0=\frac{eE}{m_e\omega_0c}$ exceeds $10^3$ for relativistically drifting particles, where $e$ the elementary charge, $m_e$ the electron static mass and $\omega_0$ the laser frequency. We will leave the details of our analytic pusher and comparison of the various pushers for a separate publication.

This paper is organized as follows: In Sec.~\ref{sect:source}, we elaborate on the origins of the first two numerical errors mentioned above. In Sec.~\ref{sect:solver}, a novel Maxwell solver amenable to finite-difference methods is proposed, which greatly improves (1) the dispersion characteristics and (2) evaluation of the transverse Lorentz force. An analysis of the dispersion relation for electromagnetic waves at all angles is provided. The Courant–Friedrichs–Lewy (CFL) stability condition and current correction for charge conservation are discussed. In Sec.~\ref{sect:simulation}, \textsc{Osiris} simulation results  based on the new solver are presented. We compare the simulation results for a single particle in a laser field in vacuum using the new solver with the standard Boris, Higuera-Cary and analytic pushers against analytic theory. The results show that the standard second-order Maxwell solver can lead to significant errors whereas the proposed solver can provide accurate results. We also compare simulation results with and without the new solver for a more collective behavior commonly referred to as direct laser acceleration (DLA). We then offer a summary and directions for future work in Sec.~\ref{sect:conclusion}. Lastly, more detailed analysis and details of the customized solver are provided in three appendices.

\section{Error sources in PIC codes}
\label{sect:source}

In this section, we provide details on the errors in the electromagnetic fields and the forces on charged particles  when using finite-difference (and some FFT-based) PIC codes. 

\subsection{Numerical dispersion}

Because of the space and time discretization of the PIC algorithm, the grid (or mesh) can be viewed as a special medium in which the electromagnetic wave is subject to a dispersion relation different than that in a vacuum. The numerical dispersion relation leads to a phase velocity that deviates from the speed of light, causing inaccuracies in the computation of particle motion. We will show later that this error is generally the largest amongst those discussed here when using a standard FDTD PIC code.

The numerical dispersion relation can be derived from the discrete Faraday's and Ampere's laws as
\begin{equation}
\label{eq:maxwell}
\text{d}_t \bm{B} = -\textbf{d}_E \times \bm{E},\quad
\text{d}_t \bm{E} = \textbf{d}_B \times \bm{B} - \bm{J},
\end{equation}
where $\text{d}_t$ and $\textbf{d}_{E,B}$ are the generalized finite-difference operators. For the remainder of the article, we use the normalized units in which $c$, $m_e$ and $e$ are equivalently viewed as unity, and the variables having time and length dimensions are normalized to reciprocals of arbitrary frequency $\omega_n$ and wavenumber $k_n\equiv \omega_n/c$. The spatial operators used in Faraday's and Ampere's laws can be different. Note that the operator used in Ampere's law should be the same as that assumed in the continuity equation for a charge-conserving scheme where Gauss's law is maintained or that used directly to solve Gauss's law. Performing a Fourier transform gives
\begin{equation}
\label{eq:maxwell_fourier}
[\omega]_t \tilde{\bm{B}} = [\bm{k}]_E \times \tilde{\bm{E}}, \quad
[\omega]_t \tilde{\bm{E}} = -[\bm{k}]_B \times \tilde{\bm{B}} - i\tilde{\bm{J}},
\end{equation}
where $[\omega]_t$ and $[\bm{k}]_{E,B}$ are the counterparts of the discrete finite-difference operators in Fourier space. By ignoring the source term $\bm{J}$, the numerical dispersion relation in vacuum can be obtained as
\begin{equation}
[\omega]_t^2-[\bm{k}]_B \cdot [\bm{k}]_E = 0.
\end{equation}
In the above derivation, Gauss's law $i[\bm{k}]_B \cdot \tilde{\bm{E}}=0$ is used. If we assume that the laser field propagates in the $\hat{1}$-direction, then its wavenumber has only a $k_1$ component and the numerical dispersion relation becomes
\begin{equation}
\label{eq:dispersion}
[\omega]_t^2-[k]_{B1}[k]_{E1}=0.
\end{equation}
In the standard leapfrog PIC algorithm, the electric field $\bm{E}$ is defined on the grid a half time step away from the magnetic field $\bm{B}$. Therefore, the operator $[\omega]_t$ has the form $[\omega]_t=\sin\left(\frac{\omega\Delta t}{2}\right)/\frac{\Delta t}{2}$, for which the phase velocity is given as
\begin{equation}
\label{eq:v_phi}
v_\phi\equiv\frac{\omega}{k_1}=\frac{2}{k_1\Delta t}\arcsin\left(\frac{\Delta t}{2}\sqrt{[k]_{B1}[k]_{E1}}\right),
\end{equation}
where $\Delta t$ is the time step and $\omega$ is the frequency of the electromagnetic wave. For the Yee mesh, where the electric and magnetic field components are stored on the staggered grid points in space as well, finite-difference operators for $[k]_{B1}$ and $[k]_{E1}$ of arbitrary order have the form
\begin{equation}
[k]_{B1,E1} = \sum_{j=1}^{p/2} C^{B,E}_j \frac{\sin\left[(2j-1)\frac{k_1\Delta x_1}{2}\right]}{\Delta x_1/2},
\end{equation}
where $p$ is the order of accuracy and $C^{B,E}_j$ is the stencil coefficient. In a conventional PIC algorithm, we usually use the same solver stencil for both Faraday's and Ampere's equations, \emph{i.e.}, $[k]_{B1}=[k]_{E1}=[k]_1$. For example, the standard Yee solver of second-order accuracy has $[k]_{B1}=[k]_{E1}=\sin(\frac{k_1\Delta x_1}{2})/\frac{\Delta x_1}{2}$.

\begin{figure}[htbp]
\centering
\includegraphics[width=\textwidth]{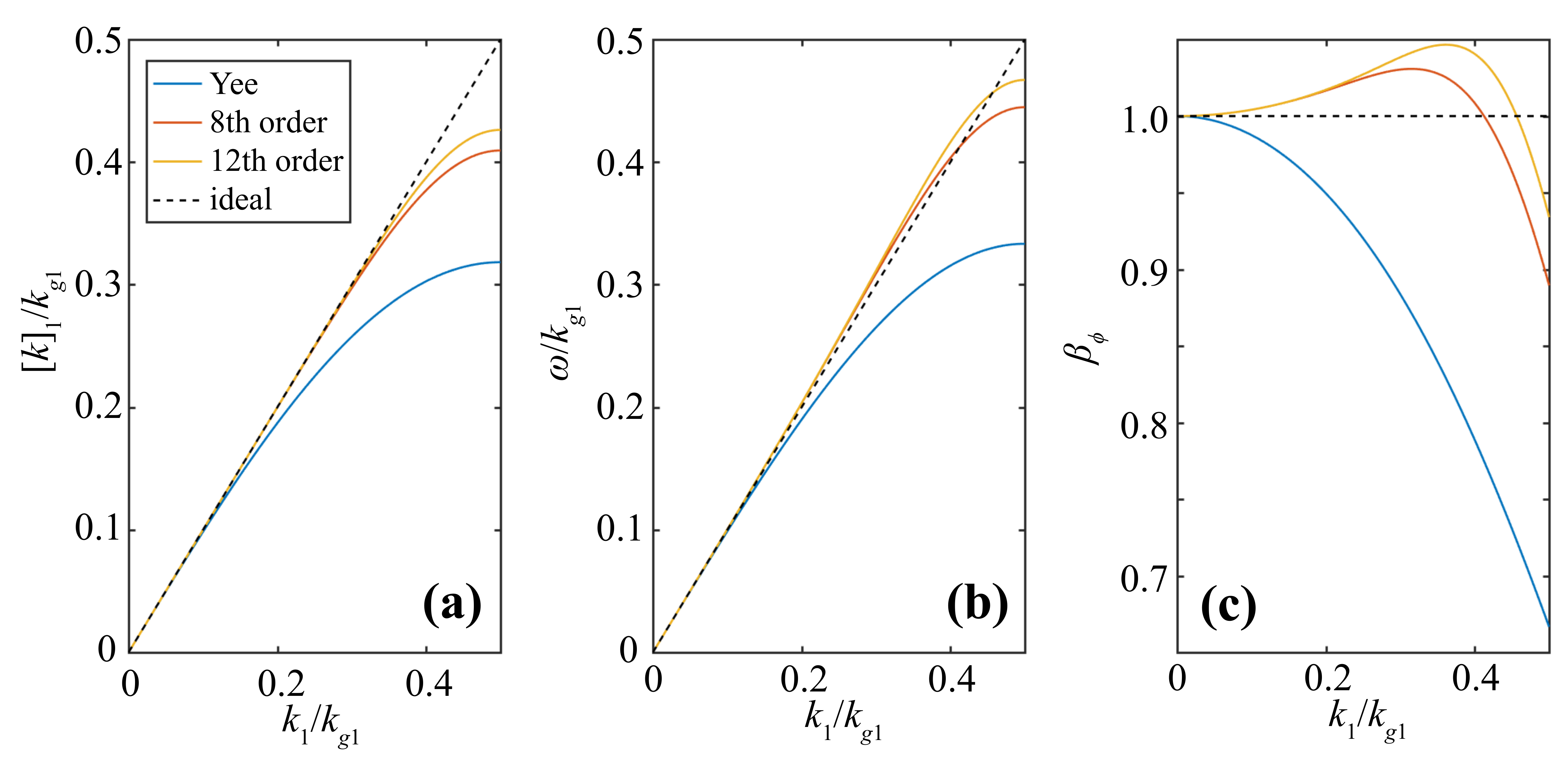}
\caption{(a)~The operator $[k]_1$ of the standard solvers with different orders of accuracy as a function of $k_1$. (b)~The numerical dispersion relation, $\omega$ vs $k_1$. (c)~The phase velocity, $\beta_\phi \equiv \omega/k_1$, as a function of $k_1$. To generate these plots, the cell size and time step were set as $\Delta x_1=0.2k_0^{-1}$ and $\Delta t=0.5\Delta x_1$. The normalization unit is defined as $k_{g1}\equiv 2\pi/\Delta x_1$.}
\label{fig:num_disp_std}
\end{figure}

Figure~\ref{fig:num_disp_std} shows the $[k]_1$, $\omega$,  and  phase velocity ($\beta_\phi \equiv \omega/k_1$) as a function of $k_1$ for finite-difference solvers of different accuracy. We can see that the dispersion relation and phase velocity of the second-order solver (Yee) can deviate significantly from real physics. Although higher-order solvers decrease the deviation, even seemingly trivial discrepancies in the phase velocity can still have cumulative effects on particle dynamics in long-duration (distance) simulations. For example, if we set the cell size to $\Delta x_1=0.2k_0^{-1}$ (5 points within a laser skin depth $k_0^{-1}$) and the time step to $\Delta t=0.5\Delta x_1$, the phase velocity of the $k_0 \approx 0.03k_{g1}$ mode is $\beta_\phi=0.99875$ (which corresponds to $\gamma_{\phi}=20$) for the second-order Yee solver. For these parameters, an ultra-relativistic particle would undergo an artificial backward phase shift of one laser wavelength after $5\times10^4$ time steps (800 laser cycles), gaining less energy than it would otherwise. However, a mildly relativistic particle with $\gamma \approx 20$ would stay in phase with a light wave moving slower than the speed of light, enabling increased energy gain from the laser. The $k_0$ mode of the 8th-order solver is superluminal, as shown in Fig.~\ref{fig:num_disp_std}(c), and the phase velocity is estimated to be $\beta_\phi\sim1.00042$. This would cause an artificial forward phase shift of one laser wavelength after $1.5\times10^5$ time steps (2,400 laser cycles) for an ultra-relativistic particle. Therefore, the numerical errors can become important for both moderately and highly relativistic particles (in long duration simulations) due to errors in the phase velocity. For modes with higher $k_1$, which are present for a light wave packet, the phase velocity deviation and artificial phase shift are more severe.

\subsection{Inaccurate calculation of the Lorentz force}

In order to illustrate how the time staggering between $\bm E$ and $\bm B$ leads to a spurious force exerted on the particles from a laser field, we start from the particle pusher used in the PIC algorithm. For simplicity, we assume the laser is polarized in the $\hat{2}$-direction (the other transverse direction being the $\hat{3}$-direction). A particle with charge $q$ is pushed according to
\begin{equation}
\label{eq:pusher}
\frac{\gamma^{n+\frac{1}{2}}\beta_2^{n+\frac{1}{2}}-\gamma^{n-\frac{1}{2}}\beta_2^{n-\frac{1}{2}}}{\Delta t} = q\left(E_2^n - \bar{\beta}_1^n \times \frac{B_3^{n-\frac{1}{2}}+B_3^{n+\frac{1}{2}}}{2} \right),
\end{equation}
where the laser fields $E_2$ and $B_3$ are interpolated from the spatial grid points and $\beta_i$ refers to the velocity of the particle in the $\hat{i}$-direction. The superscript $\langle\cdot\rangle^n$ represents the quantities at $t=n\Delta t$, and the overbar $\bar{\langle\cdot\rangle}$ represents interpolation in time. Since the magnetic field components are defined on the half time step whereas the Lorentz force is evaluated at the integer time step, $B_3$ needs to be interpolated in time. In the standard PIC algorithm, this is usually fulfilled by simply averaging, \emph{i.e.}, $\bar{B}_3^n=(B_3^{n-\frac{1}{2}}+B_3^{n+\frac{1}{2}})/2$, as shown in Eq.~(\ref{eq:pusher}). Because of the averaging, $\bar{B}_3^n$ does not equal $E_2^n$ with sufficient precision for particles moving near the speed of light in the $\hat 1$-direction, which introduces  errors when pushing the macro-particles and hence errors to each particle's trajectory. It should be noted that although the field components are also stored on spatially staggered grid points for a Yee mesh, our derivation shows that the spatial staggering has no contribution to the spurious Lorentz force.

From the discretized Maxwell equations 
in Eq.~(\ref{eq:maxwell_fourier}), the relation between $\tilde{E}_2$ and $\tilde{B}_3$ is
\begin{equation}
\label{eq:b3_e2}
\tilde{B}_3=\frac{[k]_{E1}}{[\omega]_t}\tilde{E}_2,
\end{equation}
from which it can be shown (see \ref{sect:app:force}) that in two dimensions the transverse Lorentz force exerted on the particle is
\begin{equation}
\label{eq:force2}
\frac{\tilde{F}_2(\omega,k_1,k_2)}{q}=\left[\tilde{E}_2(\omega,k_1,k_2)-\bar{\beta}_1 \tilde{B}_3(\omega,k_1,k_2)\cos\frac{\omega \Delta t}{2}\right] \tilde{S}(-k_1,-k_2),
\end{equation}
where $\tilde{S}$ is the Fourier transform of the interpolation function. The factor of $\cos\frac{\omega\Delta t}{2}$ is due to the time staggering and corresponding average in Eq.~(\ref{eq:pusher}), but the spatial staggering has no impact as aforementioned. Combining Eqs.~(\ref{eq:b3_e2}) and (\ref{eq:dispersion}), we have
\begin{equation}
\label{eq:f2_q}
\frac{\tilde{F}_2}{q}=\tilde{E}_2 \left[1-\bar{\beta}_1\sqrt{[k]_{E1}/[k]_{B1}}\cos\frac{\omega\Delta t}{2}\right] \tilde{S}.
\end{equation}
For the standard PIC algorithm with $[k]_{E1}=[k]_{B1}$, the factor $\cos\frac{\omega\Delta t}{2}$ cannot be eliminated from Eq.~(\ref{eq:f2_q}). The correct cancellation, which has the form $\tilde{F}_2=q\tilde{E}_2(1-\bar{\beta}_1)\tilde{S}$, is therefore unattainable for any solver with identical $[k]_{E1}$ and $[k]_{B1}$ operators.

\subsection{Coupling of the dispersion and Lorentz force errors}

It should be noted that the two numerical errors just described are not separable. It is possible that total error in the particle's trajectory is actually less than than that from each on their own.  This can be illustrated qualitatively by a simple case where a particle with velocity $\beta_1$ co-propagates with a monochromatic plane wave with amplitude $E_0$ and frequency $\omega_0$. The $\hat{2}$-component of the Lorentz force in the presence of the time staggering and numerical dispersion is thus $F_2^*=E_0(1-\beta_1\cos\frac{\omega_0\Delta t}{2})\cos\{[k]_0(\beta_\phi t-x_1)+\phi_0\}$, where $\beta_\phi$ is the phase velocity, $\phi_0$ is the initial phase, and $[k]_0$ is the wavenumber $k_0$ under numerical dispersion. Since the analytical force is $F_2=E_0(1-\beta_1)\cos[k_0(t-x_1)+\phi_0]$ (note $\omega_0=k_0$), it can be shown the instantaneous error in the force $\delta F_2=F_2-F_2^*$ at time $t$ is
\begin{equation}
\label{eq:df2}
\delta F_2=E_0(1-\beta_1)[\cos\phi-\cos(\phi+\delta\phi)]-2E_0\beta_1\sin^2\frac{\omega_0\Delta t}{4}\cos(\phi+\delta\phi)
\end{equation}
where $\phi\equiv k_0(t-x_1)+\phi_0$ is the analytical phase and $\delta\phi\equiv [k]_0(\beta_\phi-1)x_1$ is the phase error induced by numerical dispersion. The first term in Eq.~(\ref{eq:df2}) originates purely from the numerical dispersion while the second term couples both the dispersion and time-stagger errors together. If we use  the Yee solver as an example, the two terms are non-vanishing and $\delta\phi$ is negative. Considering a particle residing at $\phi=0$, the signs of the two terms are opposite, partially canceling the force error. The total error in $F_2$ may thus be smaller than the error from only one term. In light of this coupling between errors caused by numerical dispersion and time staggering, reducing errors from only one source might not necessarily improve overall accuracy. Therefore, finding a solution that can simultaneously reduce both the errors is of particular importance.

\section{Improved Maxwell solver}
\label{sect:solver}

\subsection{Improved dual $[k]_1$ operator}

In order to provide accurate modeling of intense laser-matter interactions, it is first important to accurately model how a single particle interacts with an intense electromagnetic wave. To achieve this using PIC codes, we need to  improve both the numerical dispersion relation and compensate for the spurious force induced by the time stagger of the $\bm E$ and $\bm B$ fields. The idea is to determine the $[k]_{E1}$ and $[k]_{B1}$ operators (in Fourier space) that minimize or eliminate errors in both the dispersion relation and Lorentz force, and then to develop finite-difference operators (in real space) that provide those desired $[k]_{E1}$ and $[k]_{B1}$ operators. If one is using an FFT-based algorithm, the desired operators can be used in Fourier space directly.

In general, we would like $[k]_{E1} = [k]_{B1} = k_1$ and $[\omega]_t =\omega$. However, dispersion errors will be minimized if the ratio $k_1^2/\omega^2$ is (at least nearly) error free, or
\begin{equation}
\frac{[k]_{E1}[k]_{B1}}{[\omega]_t^2} \rightarrow \frac{k_1^2}{\omega^2}.
\end{equation}
Furthermore, to minimize spurious terms in the Lorentz force we would like
$[k]_{E1}$ and $[k]_{B1}$ to best approximate the Lorentz force:
\begin{equation}
1-\bar{\beta}_1\sqrt{[k]_{E1}/[k]_{B1}}\cos\frac{\omega\Delta t}{2} \rightarrow 1-\bar{\beta}_1.
\end{equation}
These two conditions can be simultaneously satisfied precisely by replacing ``$\rightarrow$'' with ``$=$'', and the solution is
\begin{equation}
\label{eq:general_soln}
[k]_{E1}=\frac{[\omega]_t}{\omega\cos\frac{\omega\Delta t}{2}}k_1, \quad
[k]_{B1}=\frac{[\omega]_t\cos\frac{\omega\Delta t}{2}}{\omega}k_1.
\end{equation}
It should be pointed out that this solution is valid for any specified dispersion relation. In the context of this article, the dispersion relation of interest is that of light waves propagating along $x_1$, so we substitute the relation $\omega=k_1$ into Eq.~(\ref{eq:general_soln}) and obtain
\begin{equation}
\label{eq:new_op}
[k]_{E1}=\frac{[k]_{1,t}}{\cos\frac{k_1\Delta t}{2}}, \quad
[k]_{B1}=[k]_{1,t}\cos\frac{k_1\Delta t}{2},
\end{equation}
where $[k]_{1,t}\equiv\sin(\frac{k_1\Delta t}{2})/\frac{\Delta t}{2}$.  Note that $[k]_{1,t}$  is exactly the solver proposed by Xu~\cite{xu2019} to reduce field errors surrounding relativistic particles. For simplicity, we will call the solver associated with the $[k]_{1,t}$ operator the Xu solver for the remainder of the article. Such operators can be readily achieved by spectral (FFT) based solvers, but are impossible to be matched exactly by standard finite-difference solvers. To approximate these operators using  a finite-difference method in broad regions of $k_1$ space, we follow the methodology in Ref.~\cite{li2017}.  The target forms for $[k]_{B1}$ and $[k]_{E1}$ are achieved by extending the solver stencil and customizing its coefficients. The number of stencil coefficients is increased from $p/2$ to arbitrary $M$, where $M>p/2$. The detailed method of fitting the $[k]_{E1}$ and $[k]_{B1}$ operators using customized coefficients is described in  \ref{sect:app:solver}.

\begin{figure}[htbp]
\centering
\includegraphics[width=\textwidth]{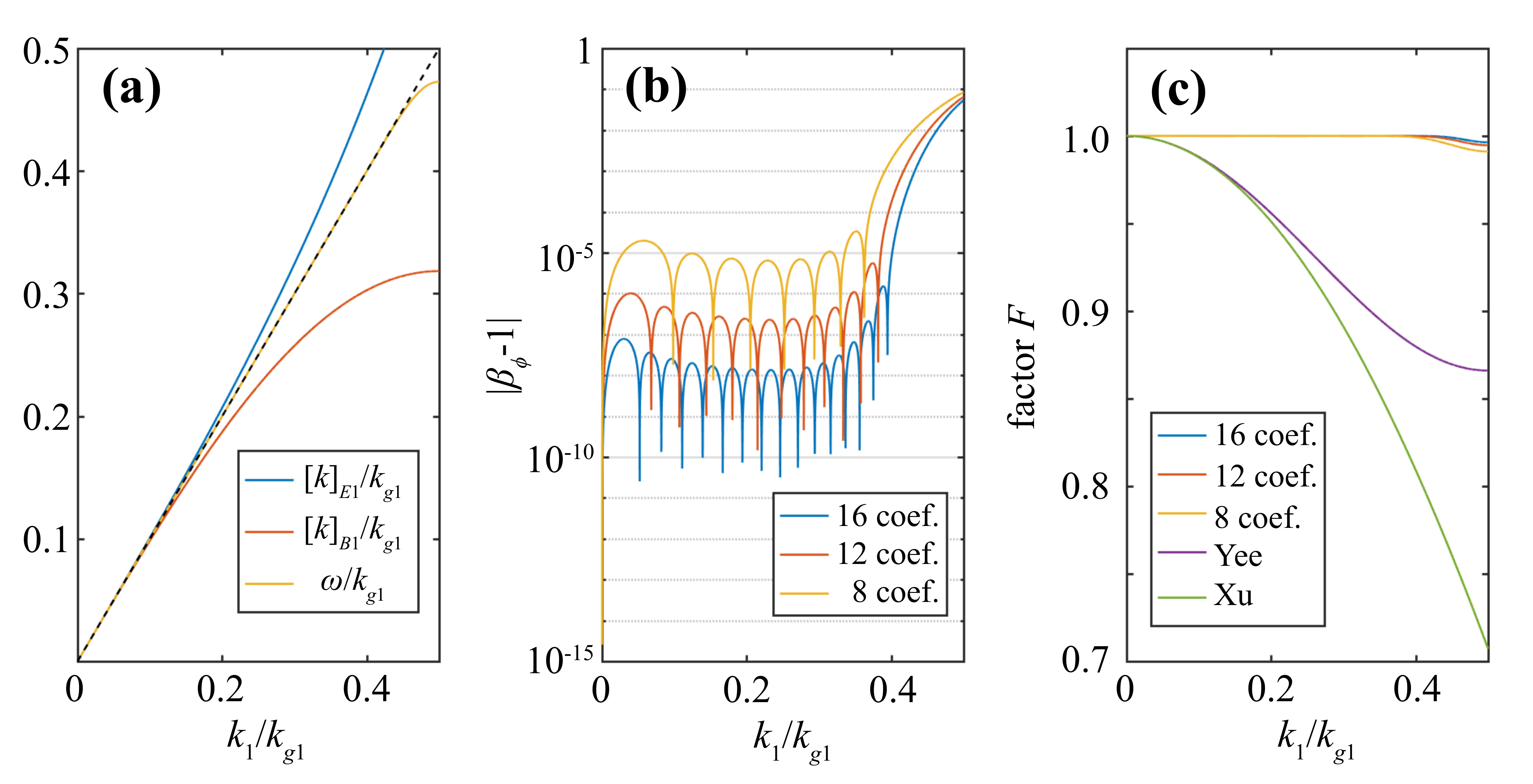}
\caption{(a) Calculated $[k]_{E1}$ and $[k]_{B1}$ operators fitted by the coefficient customization method (using a 16-coefficient stencil). The yellow line represents the numerical dispersion relation, $[\omega]_t=\sqrt{[k]_{E1}[k]_{B1}}$. (b) Error in phase velocity for solver stencils with varying number of coefficients. (c) Comparison of the cancellation factor $F\equiv\sqrt{[k]_{E1}/[k]_{B1}}\cos\frac{\omega\Delta t}{2}$ (ideally $F=1$) between the proposed solvers with different stencils, the standard Yee solver and the Xu solver. The numerical parameters are $\Delta x_1=0.2k_0^{-1}$, $\Delta t=0.5\Delta x_1$ and $k_{g1}\equiv 2\pi/\Delta x_1$.}
\label{fig:num_disp_improved}
\end{figure}

In Figure~\ref{fig:num_disp_improved} we present results for the $k_1$-space operators, the numerical dispersion errors and the Lorentz force errors for a 16-coefficient customized stencil with $\Delta x_1=0.2k_0^{-1}$ and $\Delta t=0.5\Delta x_1$. In Fig.~\ref{fig:num_disp_improved}(a) the $[k]_{E1}$ (blue line) and $[k]_{B1}$ (red line) operators are shown as functions of $k_1$. Although they seem to deviate more from their individual ideal forms than do the standard higher-order solvers [see Fig.~\ref{fig:num_disp_std}(a)], the resulting numerical dispersion relation $[\omega]_t=\sqrt{[k]_{E1}[k]_{B1}}$ denoted by the yellow line is clearly better than for the standard operators [see Fig.~\ref{fig:num_disp_std}(b)]. In Fig.~\ref{fig:num_disp_improved}(b), we compare the phase velocity errors when using $[k]_{E1}$ and $[k]_{B1}$ fitted with different stencil widths. It can be seen that within the range $0<k_1<0.3k_{g1}$, a negligible phase velocity error ($\sim10^{-5}$) is achieved using only 8 stencil coefficients. Since the high-$k$ modes [$k_1>0.3k_{g1}$ in Fig.~\ref{fig:num_disp_improved}(b)] with relatively large phase velocity errors can usually be filtered out as they lie outside the Fourier modes of physical importance, such a result is good enough for most cases. The comparison in Fig.~\ref{fig:num_disp_improved}(b) shows that the dispersion relation can be further improved by using solvers with wider stencils. For example, increasing the number of stencil coefficients from 8 to 16 improves the accuracy of the phase velocity by nearly two orders of magnitude.

In Fig.~\ref{fig:num_disp_improved}(c), we compare the errors in the Lorentz force for a plane wave with different solvers, as defined by the cancellation factor $F\equiv\sqrt{[k]_{E1}/[k]_{B1}}\cos\frac{\omega\Delta t}{2}$ [see Eq.~(\ref{eq:f2_q})], where $\omega$ is calculated under numerical dispersion. In the continuous limit, this factor should be unity, $F=1$. For the standard Yee solver and any others with $[k]_{B1}=[k]_{E1}$, the $F$ factor has a noticeable deviation in almost the entire first Brillouin zone, $k_1< k_{g1}$. Even for the Xu solver (green line), which exhibits the correct dispersion relation, we still have a very large deviation in the $F$ factor. However, the proposed solvers with different stencil widths significantly improve the $F$ factor. Within a considerably wide range of $k_1$, the cancellation factors are very close to 1, as seen in Fig.~\ref{fig:num_disp_improved}(c). The very high $k$ modes will be filtered out as mentioned before; the proposed solver thus provides an improved dispersion relation and field cancellation for the $k_1$ range of interest. 

\subsubsection{Obliquely traveling waves}

Although the proposed solver is designed for electromagnetic waves propagating parallel to the $\hat{1}$-direction, its numerical dispersion relation is still better than that of the Yee solver even when the incident wave travels at a small angle. The errors with the proposed solver gradually increase with increasing angle until they are identical to those with the Yee solver for propagation at $\pi/2$ with respect to the $\hat{1}$-direction (maintaining consistent time step and cell sizes). To illustrate this feature, we assume an obliquely incident plane wave with wave vector $\bm{k}_0$ traveling in the $x_1$-$x_2$ plane ($k_3=0$). Let the incident angle be $\theta$, so that $k_1=k_0\cos\theta$ and $k_2=k_0\sin\theta$. Since $k_2$ is now non-vanishing, we need to include $[k]_{E2}$ and $[k]_{B2}$ operators when calculating the phase velocity using Eq.~(\ref{eq:v_phi}). Here, for both the Yee and proposed solvers, the operators in $x_2$ have identical forms, $[k]_{E2}=[k]_{B2}=\sin\left(\frac{k_2\Delta x_2}{2}\right)/\frac{\Delta x_2}{2}$. Using the operators defined in Eq.~(\ref{eq:new_op}) for the proposed solver and $[k]_{E1}=[k]_{B1}=\sin\left(\frac{k_1\Delta x_1}{2}\right)/\frac{\Delta x_1}{2}$ for the Yee solver, we can write the phase velocity in a unified form,
\begin{equation}
\label{eq:vphi_ob}
v_{\phi}=\frac{2}{k_0\Delta t}\arcsin\left[\frac{\Delta t}{2}\sqrt{\left(\frac{\sin(\eta k_0\Delta t\cos\theta/2)}{\eta\Delta t/2}\right)^2+\left(\frac{\sin(k_0\Delta x_2\sin\theta/2)}{\Delta x_2/2}\right)^2}\right],
\end{equation}
where $\eta=1$ and $\eta=\frac{\Delta x_1}{\Delta t}$ for the proposed and Yee solvers, respectively. Performing a Taylor expansion in $\theta$ gives
\begin{align}
\begin{split}
v_{\phi}=&\frac{2}{k_0\Delta t}\arcsin\left[\eta^{-1}\sin\left(\frac{\eta k_0\Delta t}{2}\right)\right]\\
&-\frac{\eta[\sin(\eta k_0\Delta t)-\eta k_0\Delta t]}{2\sqrt{(1-\cos(\eta k_0\Delta t))(2\eta^2-1+\cos(\eta k_0\Delta t))}}\theta^2 + O(\theta^3).
\end{split}
\end{align}

\begin{figure}[htbp]
\centering
\includegraphics[width=0.9\textwidth]{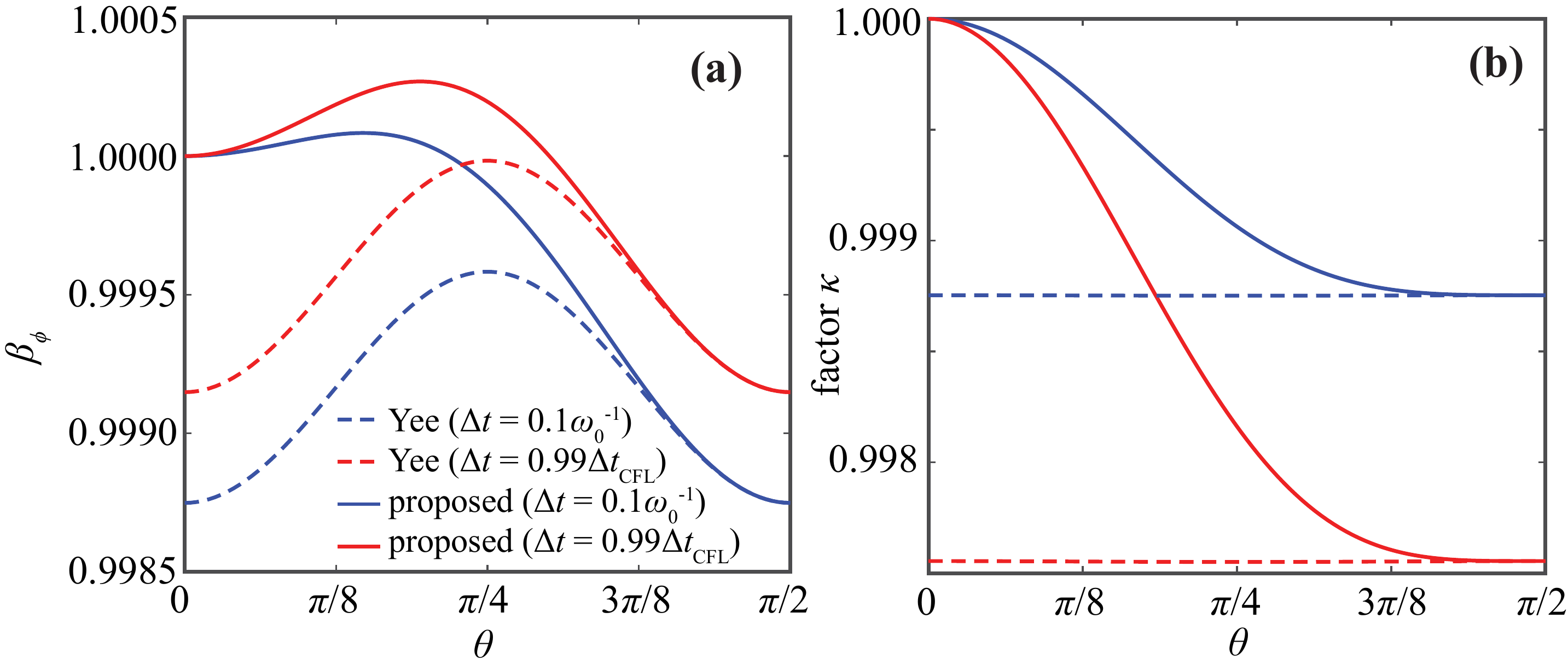}
\caption{(a) Phase velocity, $\beta_\phi$, and (b) factor $\kappa$ (see text for definition) vs incident angle. To generate these plots, we take $\Delta x_1=\Delta x_2=0.2k_0^{-1}$. Of particular note is, in (b), the curves for the Yee solver (dashed lines) actually have small-amplitude modulations and thus cannot be considered independent of $\theta$.}
\label{fig:oblique_inc}
\end{figure}

The leading term for the proposed solver  exactly equals to unity ($\eta=1$), while for the Yee solver where $\eta=\frac{\Delta x_1}{\Delta t}>1$ due to the CFL stability condition, the leading term is always less than 1. Since the coefficient of the $\theta^2$ term is positive, waves moving at a small incident angle will travel slightly faster than those parallel to the $\hat{1}$-direction. Figure~\ref{fig:oblique_inc}(a) shows the phase velocity as a function of $\theta$ according to Eq.~(\ref{eq:vphi_ob}) for a reasonable choice of $k_0$ with $\Delta x_1=\Delta x_2=0.2k_0^{-1}$. For $\Delta t=0.1\omega_0^{-1}$ (blue lines), much smaller than the Courant limit $\Delta t_\text{CFL}=2^{-1/2}\omega_0^{-1}$, the proposed solver always has the smallest errors in the phase velocity (closer to the speed of light) for all angles less than $\pi/2$. On the other hand, for $\Delta t\simeq\Delta t_\text{CFL}$ (red lines), the proposed solver still has smaller errors in the phase velocity for a wide range of $\theta$. As is well known, we can see that for the Yee solver with $\Delta t\simeq\Delta t_\text{CFL}$, there is an angle for which the phase velocity is exactly equal to unity (for this case with square cells the angle is $\pi/4$). For angles less than $\sim \pi/8$, it is clear that for both values of $\Delta t$ the proposed solver has smaller errors in the phase velocity. It can also be seen that for a given time step, the dispersion errors for the proposed solver converge to those for the Yee solver (while remaining slightly smaller) as the angle approaches $\pi/2$. This, together with the fact that the phase velocity at small angles for the proposed solver---unlike the Yee solver---monotonically converges to unity as $\Delta t$ is reduced, allows for convergence tests by reducing the time step (since the field solver and the pusher both get more accurate).

We can also consider the accuracy of the Lorentz force in the $\hat{2}$-direction for a plane wave moving at an angle $\theta$ and a particle moving in the $\hat{1}$-direction. A similar analysis could be done for the force in the $\hat{1}$-direction. The $\hat{2}$-component of the Lorentz force has the same form as described by Eq.~(\ref{eq:force2}), but due to the laser moving at an angle there is also a component of $\bm k$ in the $\hat{2}$-direction, making $\tilde{B}_3$ the sum of two terms, 
\begin{equation}
\tilde{B}_3=\frac{[k]_{E1}}{[\omega]_t}\tilde{E}_2-\frac{[k]_{E2}}{[\omega]_t}\tilde{E}_1,
\end{equation}
according to Eq.~(\ref{eq:maxwell_fourier}). Substituting into Eq.~(\ref{eq:force2}), we have
\begin{align}
\begin{split}
\frac{\tilde{F}_2}{q}=&\tilde{E}_2\left[1-\bar{\beta}_1\frac{[k]_{E1}}{\sqrt{[k]_{E1}[k]_{B1}+[k]_{E2}[k]_{B2}}}\cos\left(\frac{\omega\Delta t}{2}\right)\right]\tilde{S}\\
&+\tilde{E}_1\bar{\beta}_1\frac{[k]_{E2}}{\sqrt{[k]_{E1}[k]_{B1}+[k]_{E2}[k]_{B2}}}\cos\left(\frac{\omega\Delta t}{2}\right)\tilde{S} \\
=& \tilde{E}_0\left[\cos\theta-\bar{\beta}_1\frac{[k]_{E1}\cos\theta+[k]_{E2}\sin\theta}{\sqrt{[k]_{E1}[k]_{B1}+[k]_{E2}[k]_{B2}}}\cos\left(\frac{\omega\Delta t}{2}\right)\right]\tilde{S},
\end{split}
\end{align}
where $\tilde{E}_0$ is the complex amplitude of the plane wave and we have used $\tilde{E}_2=\tilde{E}_0\cos\theta$ and $\tilde{E}_1=-\tilde{E}_0\sin\theta$ to simplify the expression. The term introducing numerical errors is the factor $\kappa=\frac{[k]_{E1}\cos\theta+[k]_{E2}\sin\theta}{\sqrt{[k]_{E1}[k]_{B1}+[k]_{E2}[k]_{B2}}}\cos\frac{\omega\Delta t}{2}$, which reduces to $\kappa = \frac{k_1 \cos\theta + k_2 \sin\theta}{k_0} = 1$ in the continuous limit. In Fig.~\ref{fig:oblique_inc}(b) we plot $\kappa$ as a function of $\theta$ for the $[\bm{k}]_{E,B}$ operators corresponding to the Yee and proposed solvers. It can be seen that for the Yee solver the error is nearly constant over all angles and gets smaller as $\Delta t$ is reduced. On the other hand, the factor is unity for the proposed solver at an angle of 0 and is always closer to unity (for all angles) than for the corresponding Yee solver.

\subsubsection{Behavior in a plasma}

The proposed solver also gives  more accurate  dispersion relation for light in a plasma. In a cold and static plasma with the ions assumed to be immobile, it can be shown (see \ref{sect:app:nd}) that the simple numerical dispersion relation
\begin{equation}
\label{eq:plasma_nd0}
\left([\omega]_t^2-[\bm{k}]_E\cdot[\bm{k}]_B+s(\omega,\bm{k})\omega_p^2\right) 
\left([\omega]_t^2-s(\omega,\bm{k})\omega_p^2\right) = 0
\end{equation}
is satisfied when using a momentum-conserving scheme where the interpolation function for the electric field is identical to the deposition function for the charge (current). Here, $s$ is an auxiliary parameter related to the interpolation function and the aliasing effect, as defined in \ref{sect:app:nd}. After inspection, it can be seen that the first term in the above equation corresponds to the electromagnetic mode, while the second term corresponds to the Langmuir mode.

\begin{figure}[htbp]
\centering
\includegraphics[width=0.8\textwidth]{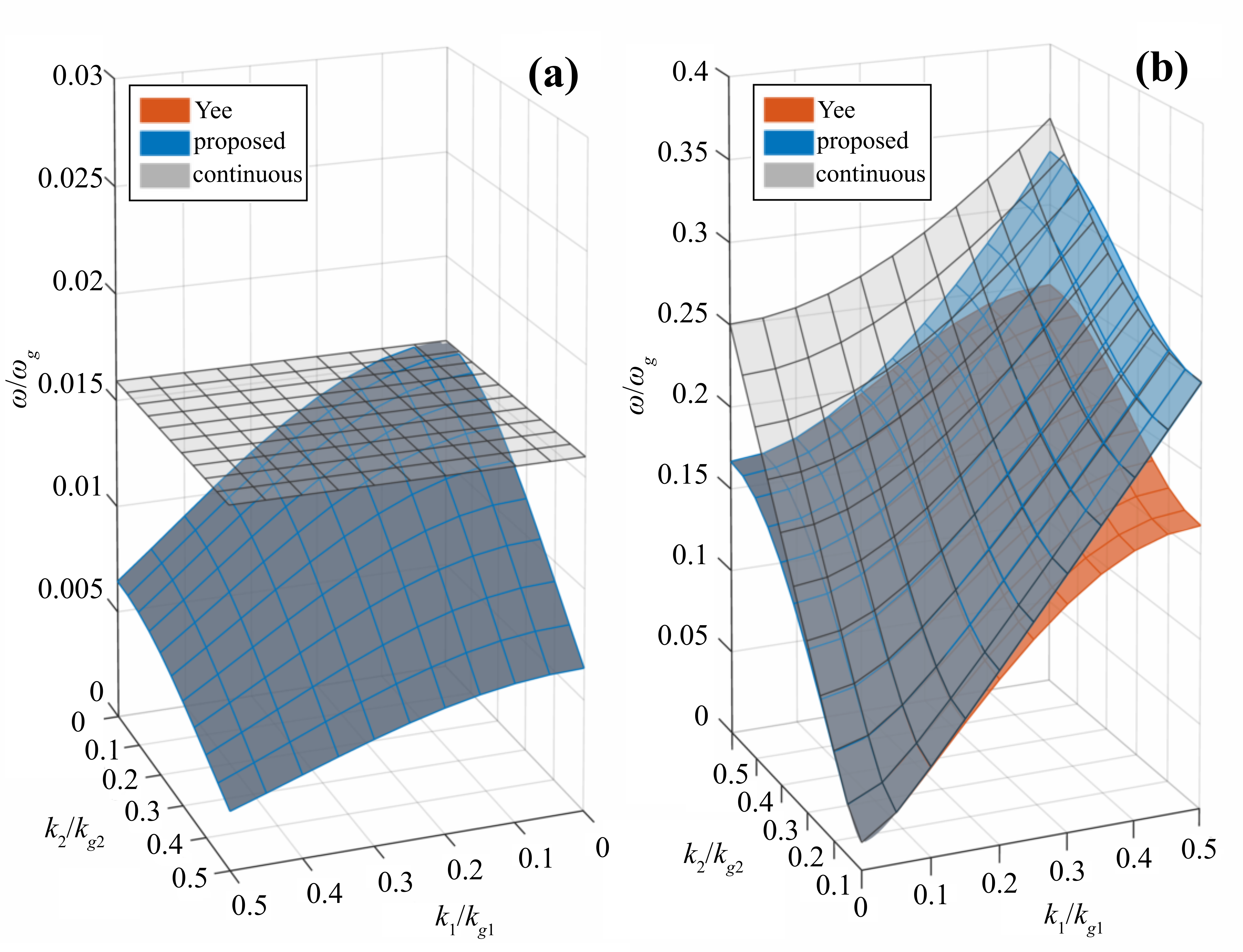}
\caption{Comparison of the $\omega$-$\bm{k}$ relation between the different solvers and theory for the (a) Langmuir mode and (b) electromagnetic mode as defined in Eq.~(\ref{eq:plasma_nd0}). To generate these plots, we choose $\omega_n=\omega_p=1$, $k_n=\omega_n/c=1$, $\Delta x_1=\Delta x_2=0.2$ and $\Delta t=0.5\Delta x_1$; note $k_{g1}\equiv 2\pi/\Delta x_1$. For simplicity, we neglect aliasing, \emph{i.e.}, only keeping the $(\mu,\bm{\nu})=0$ term in the summation sign of $s(\omega,\bm{k})$ (see definition in \ref{sect:app:nd}). Note that the positions of coordinate zero are different in (a) and (b) for better visibility.}
\label{fig:plasma_nd}
\end{figure}

In Fig.~\ref{fig:plasma_nd} we plot the numerical dispersion relations for both modes using the Yee and proposed solvers and compare them against continuous-limit expressions for $\omega_p=1$. To generate these plots, we assume the aliasing effect is negligible and only solve Eq.~(\ref{eq:plasma_nd0}) in the first quadrant of the fundamental Brillouin zone, \emph{i.e.}, $(k_1,k_2)\in [0,k_{g1}/2]\times[0,k_{g2}/2]$, where $k_{gi}\equiv\frac{2\pi}{\Delta x_i}$. The Langmuir mode is shown in Fig.~\ref{fig:plasma_nd}(a), where there is no observed difference between the Yee and proposed solvers because the numerical dispersion relation $[\omega]_t^2=s(\omega,\bm{k})\omega_p^2$ does not explicitly rely on the $[\bm{k}]_{E,B}$ operators. In both cases, the dispersion relation depends on $\bm{k}$ because of the interpolation function. The value of $\omega$ is $\omega_p$ for $\bm{k}=0$, and then $\omega$ decreases as the magnitude of $\bm{k}$ increases.

Figure~\ref{fig:plasma_nd}(b) shows the $\omega$-$\bm{k}$ relation for the electromagnetic mode. It can be seen that the Yee-solver surface resides well below the continuous-limit result, while that of the proposed solver falls in-between the two. When $k_2=0$ the surface of the proposed solver converges to the continuous limit, whereas when $k_1=0$ it converges to the curve of the Yee solver. Therefore, even though the proposed solver was specifically designed to optimize the behavior of single particles interacting with electromagnetic waves in vacuum, the proposed solver still gives a significantly more accurate dispersion relation than does the standard Yee solver for electromagnetic waves in a plasma. For waves propagating roughly along the $\hat{1}$-direction in a cold plasma, the proposed solver is nearly as accurate as it is in vacuum. We have only plotted results for square cells, but the general conclusions still hold for rectangular cells.

\subsection{CFL stability condition}

The use of different finite-difference  stencils in Ampere's and Faraday's laws leads to a different Courant–Friedrichs–Lewy (CFL) stability condition than that obtained for a conventional solver. According to the numerical dispersion relation in Eq.~(\ref{eq:dispersion}), the following constraint on the time step is obtained in order that $\omega$ be a real number for a real wave number:
\begin{equation}
\frac{\Delta t}{2}\sqrt{\left(\sum_{j=1}^M C_j^B s_{x_1,j}\right) \left(\sum_{j=1}^M C_j^E s_{x_1,j}\right) + s_{x_2,1}^2 } \leq 1,
\end{equation}
where $s_{x_i,j}=\sin[(2j-1)k_i\Delta x_i/2]/(\Delta x_i/2)$. We point out that the Yee operator for the $\hat{2}$-direction is included in the above inequality for the general 2D scenario. Noting that $|s_{x_i,j}|\leq\frac{2}{\Delta x_i}$, it can be shown that a sufficient condition (CFL limit for $\Delta t$) for the above inequality is
\begin{equation}
\Delta t\leq 1\left/ \sqrt{\left(\sum_{j=1}^M|C_j^B|\right) \left(\sum_{j=1}^M|C_j^E|\right) \frac{1}{\Delta x_1^2} + \frac{1}{\Delta x_2^2}} \right. .
\end{equation}

\subsection{Current correction for charge conservation}

In a typical FDTD PIC code, the electromagnetic fields are advanced via Faraday's and Ampere's laws, while Gauss's law is maintained by applying a charge-conserving current deposition scheme similar to that in Ref.~\cite{esirkepov2001}. The referenced deposition scheme is second-order-accurate in all directions, which means that Gauss's law is satisfied exactly for the standard second-order Yee solver. However, when using the proposed solver with a modified stencil in the $\hat{1}$-direction, the existing current deposition can no longer be charge conserving without a corresponding current correction. We next show that if we modify the second-order-accurate current in the $\hat{1}$-direction in Fourier space (performing an FFT only along the $\hat{1}$-direction) as follows,
\begin{equation}
\label{eq:correction}
\tilde{J}_{c,1}=\frac{[k]_{B1}}{[k]_{1,\text{Yee}}}\tilde{J}_1,
\end{equation}
that the continuity equation and hence Gauss's law are satisfied for the modified stencil, where $[k]_{1,\text{Yee}}=\sin(\frac{k_1\Delta x_1}{2})/\frac{\Delta x_1}{2}$ is the operator corresponding to the standard second-order Yee solver.

The existing second-order-accurate charge-conserving current deposition satisfies the following finite-difference representation of the continuity equation:
\begin{equation}
\text{d}_t\rho+\textbf{d}_\text{Yee} \cdot \bm{J}^{n+\frac{1}{2}}=0,
\end{equation}
where $\text{d}$ refers to differential finite-difference operators. Performing a Fourier transform in the $\hat{1}$-direction and using the corrected current from Eq.~(\ref{eq:correction}), we have
\begin{align}
\begin{split}
\text{d}_t\tilde{\rho}+i[k]_{1,\text{Yee}}\tilde{J}_{c,1}+i[k]_{2,\text{Yee}}\tilde{J}_2 &= \\
\text{d}_t\tilde{\rho}+i[k]_{B1}\tilde{J}_{1}+i[k]_{2,\text{Yee}}\tilde{J}_2 &= 0.
\end{split}
\end{align}
Combining this with the divergence of Ampere's law yields
\begin{equation}
\text{d}_t\left(\tilde{\rho}+i[k]_{B1}\tilde{E}_1+i[k]_{2,\text{Yee}}\tilde{E}_2\right)=0,
\end{equation}
which indicates that Gauss's law $i[\bm{k}]_B\cdot\tilde{\bm{E}}=-\tilde{\rho}$ is satisfied for later times if it is satisfied at $t=0$, where $[\bm{k}]_B=([k]_{B1},[k]_{2,\text{Yee}})$.

\section{Sample simulations}
\label{sect:simulation}

In this section, we present two examples where the proposed solver improves results from simulations. A single relativistic charged particle co-propagating with a laser pulse is simulated in the first example. The second is a more complicated scenario, where an electron bunch is injected and accelerated directly by the wakefield and laser pulse in a laser wakefield accelerator. We used the PIC code \textsc{Osiris}~\cite{fonseca2002, hemker2015}, where the proposed algorithm has been implemented. 

For a comparative study, we will show not only the results of the proposed solver, but those of all solvers listed below:
\begin{enumerate}
\item Standard Yee solver.
\item Xu solver with $[k]_{B1}=[k]_{E1}=[k]_{1,t}$ (see Ref.~\cite{xu2019}). This solver is considered to have a good dispersion relation, but because it uses identical $[k]_{E1}$ and $[k]_{B1}$ it does not correct for the time-stagger errors in the magnetic field.
\item Yee solver with field time-stagger correction (Yee t-stagger). As aforementioned, any dispersion relation can be set as the objective in Eq.~(\ref{eq:general_soln}), not just $\omega=k_1$. By letting $[\omega]_t=[k]_{1,\text{Yee}}$, the solver retains the dispersion errors of the Yee solver while possessing the time-stagger correction in the transverse force.
\item Proposed solver, with a good dispersion relation and field time-stagger correction. This solver can be viewed as the time-stagger-corrected version of the Xu solver.
\end{enumerate}

The purpose of doing the comparison is to demonstrate that both dispersion and time-staggering errors can contribute significant numerical errors to the motion of a single particle in an intense laser and in wakefields, and that correcting one without the other can actually make the errors larger in some cases. Therefore, correcting both numerical artifacts is important. In these comparisons we also use various particle pushers as described below. We emphasize that the results can depend on the choices of the cell size, the aspect ratio of the cells for multi-dimensional cases and the time step, in addition to the solver and pusher. These examples are not intended to be exhaustive, but illustrative.

\subsection{Single particle in a laser field}
\label{sect:simulation:single_part}

\begin{figure}[htbp]
\centering
\includegraphics[width=0.9\textwidth]{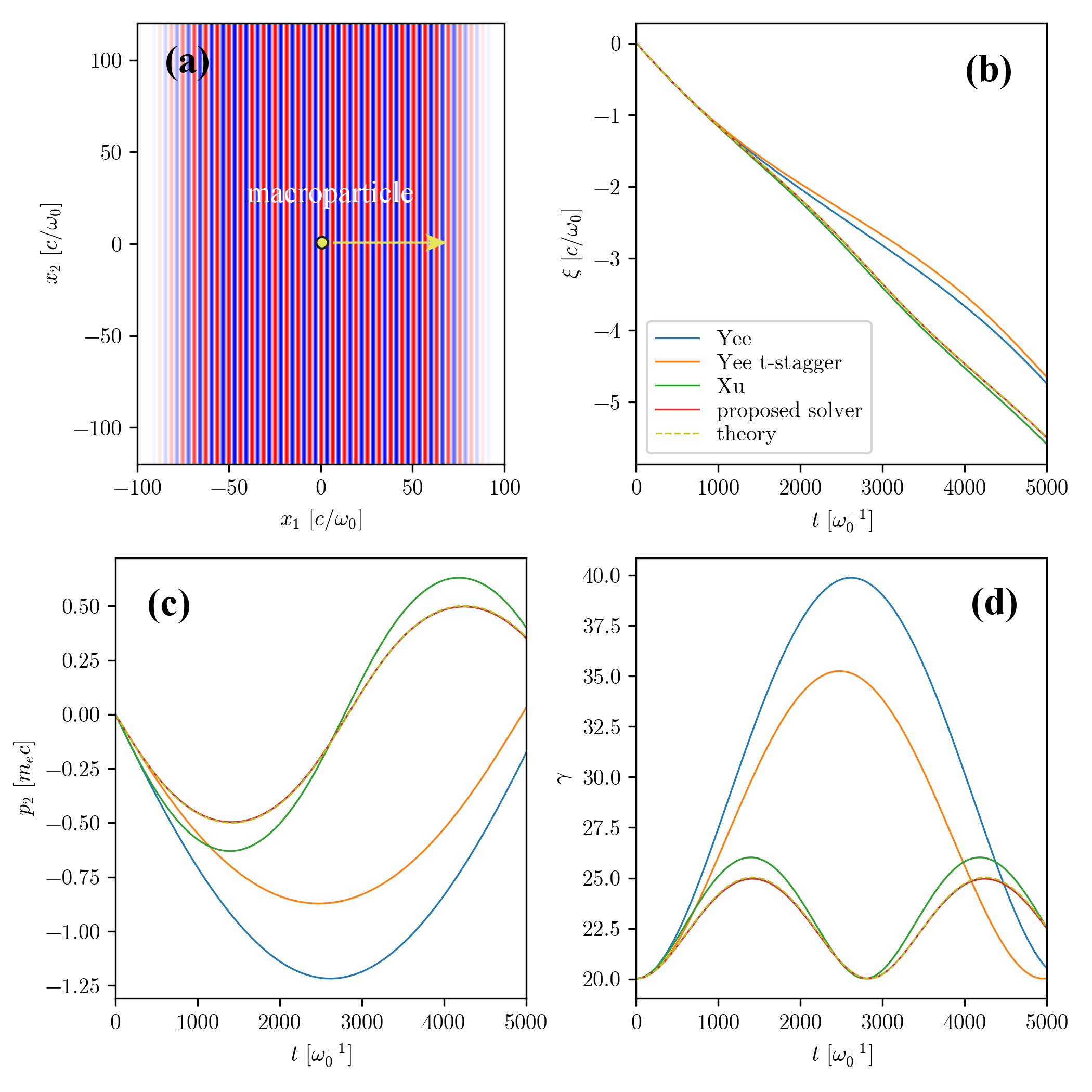}
\caption{Single-particle trajectories with an initial drift of $\gamma_0=20$ in a laser field of amplitude $a_0=0.5$. (a)~Initial configuration of the plane-wave electric field and macro-particle position. Evolution of the macro-particle (b)~ phase, $\xi=x_1-t$, (c)~transverse momentum and (d)~Lorentz factor for different solvers. The Higuera-Cary pusher is used in all the simulations.}
\label{fig:sp1}
\end{figure}

In the first set of 2D test simulations, we initialized a single relativistic macro-particle which co-propagates with a plane-wave laser pulse polarized in the $\hat{2}$-direction. Figure~\ref{fig:sp1}(a) shows the initial configuration of the simulation, where moving-window (periodic) boundaries are used in the $\hat{1}$-direction ($\hat{2}$-direction). The pattern colored by red and blue represents the $E_2$ component of the laser pulse. For simplicity, the laser is a plane wave and has a super-Gaussian longitudinal profile with a 100 $k_0^{-1}$ long flat-top. Thus the laser field has no diffraction as it propagates, and the particle always feels a constant laser amplitude. In this section, the time step ($\Delta t=0.05\omega_0^{-1}$) and cell sizes ($\Delta x_1=0.2k_0^{-1}, \Delta x_2=20 k_0^{-1}$) are fixed for all the simulations. The theoretical results plotted in each figure (dashed lines) are calculated using analytic solutions, \emph{e.g.}, see Ref.~\cite{yang2011}.

In the first example, a drifting particle with $\gamma_0=20$ is initialized inside laser fields of moderate amplitude ($a_0=0.5$) at a location where the laser electric field (vector potential) is at a maximum (zero). In Figs.~\ref{fig:sp1}(b)-(d), the particle trajectories are compared between the above-listed solvers used in conjunction with the Higuera-Cary (HC) pusher~\cite{higuera2017}. We have found that the HC pusher is generally better than the standard Boris: it exhibits the advantages of the Vay pusher~\cite{vay2008} for relativistically drifting particles without issues for non-relativistic particles.  We leave a more detailed comparison of the choice in pushers for a later publication.

Figure~\ref{fig:sp1}(b) shows the change of particle phase, $\xi=x_1-t$. Since the use of the Yee-type solvers (types 1 and 3) leads to the laser fields traveling slower than the speed of light (but still faster than the particle), the test particle undergoes a significantly smaller dephasing than for the Xu and proposed solvers. This artifact in the dispersion relation is also reflected in the oscillation period of the transverse momentum, $p_2$, as shown in Fig.~\ref{fig:sp1}(c); the Yee-type solvers have much larger oscillation periods than the others, while the solvers with a corrected dispersion relation (Xu and proposed) have similar periods that agree well with the theoretical prediction.

The adverse impact induced by the time staggering is primarily manifested in the oscillation amplitude; the amplitudes of both the standard Yee and Xu solvers in Fig.~\ref{fig:sp1}(c) are larger than their counterparts that have the time-stagger correction. As is well known, $p_2$ satisfies the canonical momentum conservation,
\begin{equation}
p_2-a_L =\text{const.},
\end{equation}
where $a_L$ is the normalized vector potential of the laser pulse. Since the test particle is initially stationary in the $\hat{2}$-direction and placed where $a_L=0$, the subsequent evolution of $p_2$ is subject to $p_2=a_L$. As the test particle progressively dephases,  $p_2$ should oscillate between $-a_0$ and $a_0$. In this regard, only the proposed solver gives a convincing solution. Figure~\ref{fig:sp1}(d) shows the change in $\gamma$, where except for the proposed solver, all others overestimate the energy gain to different degrees. It is worth noting that we have also done comparisons with the standard Boris pusher, and the results are almost identical to those with the HC pusher, implying that the discretization error on particle velocity has little impact on such problems with relatively low $a_0$.

\begin{figure}[htbp]
\centering
\includegraphics[width=0.9\textwidth]{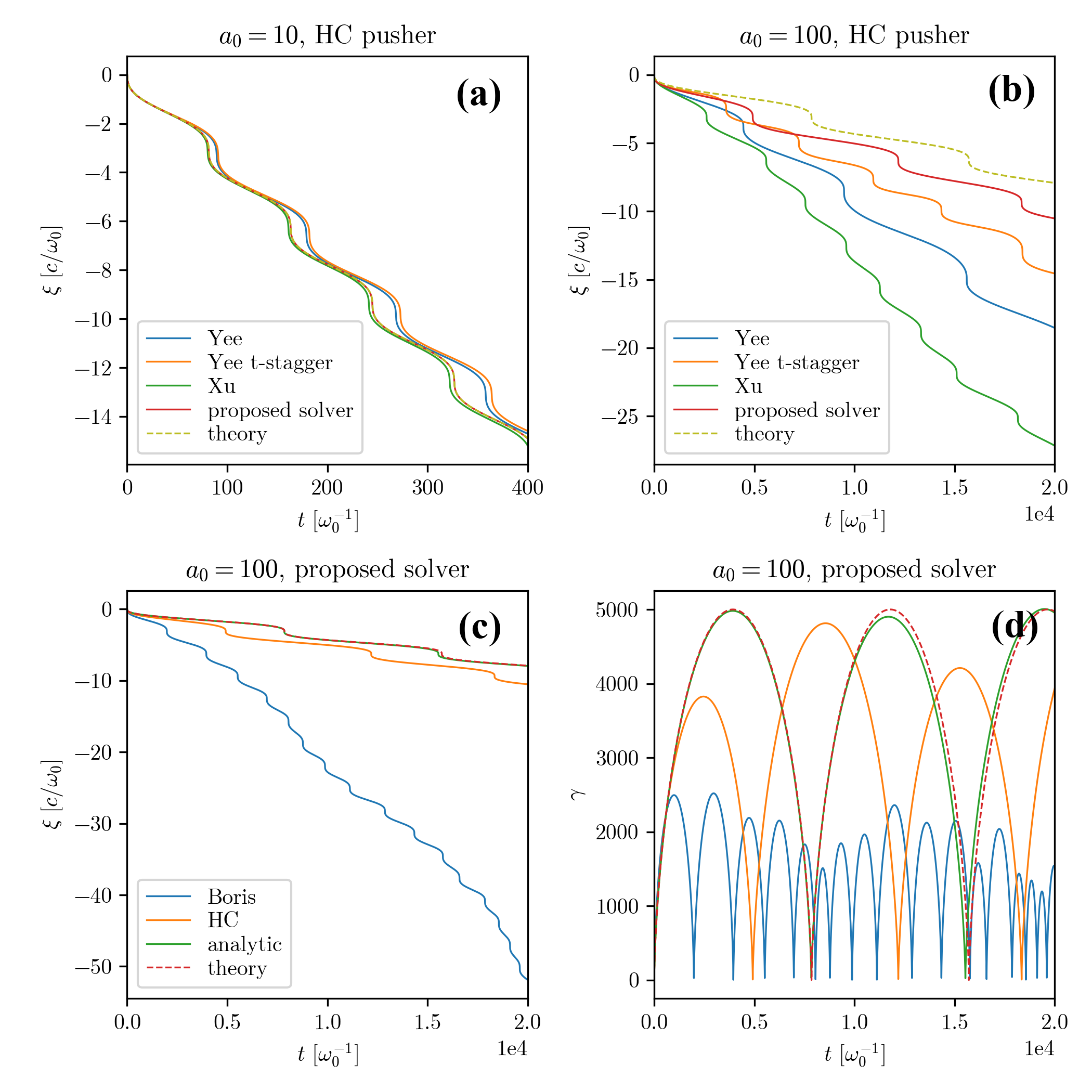}
\caption{Single-particle motion for a particle starting at rest in laser fields of varying amplitude. Evolution of the particle phase, $\xi=x_1-t$, for (a) $a_0=10$ and (b) $a_0=100$ using the HC pusher with various solvers. The evolution of the (c)~phase, $\xi$, and (d)~Lorentz factor of a particle in the presence of a laser field with $a_0=100$ using the proposed solver combined with different particle pushers.}
\label{fig:sp2}
\end{figure}

\begin{figure}[htbp]
\centering
\includegraphics[width=\textwidth]{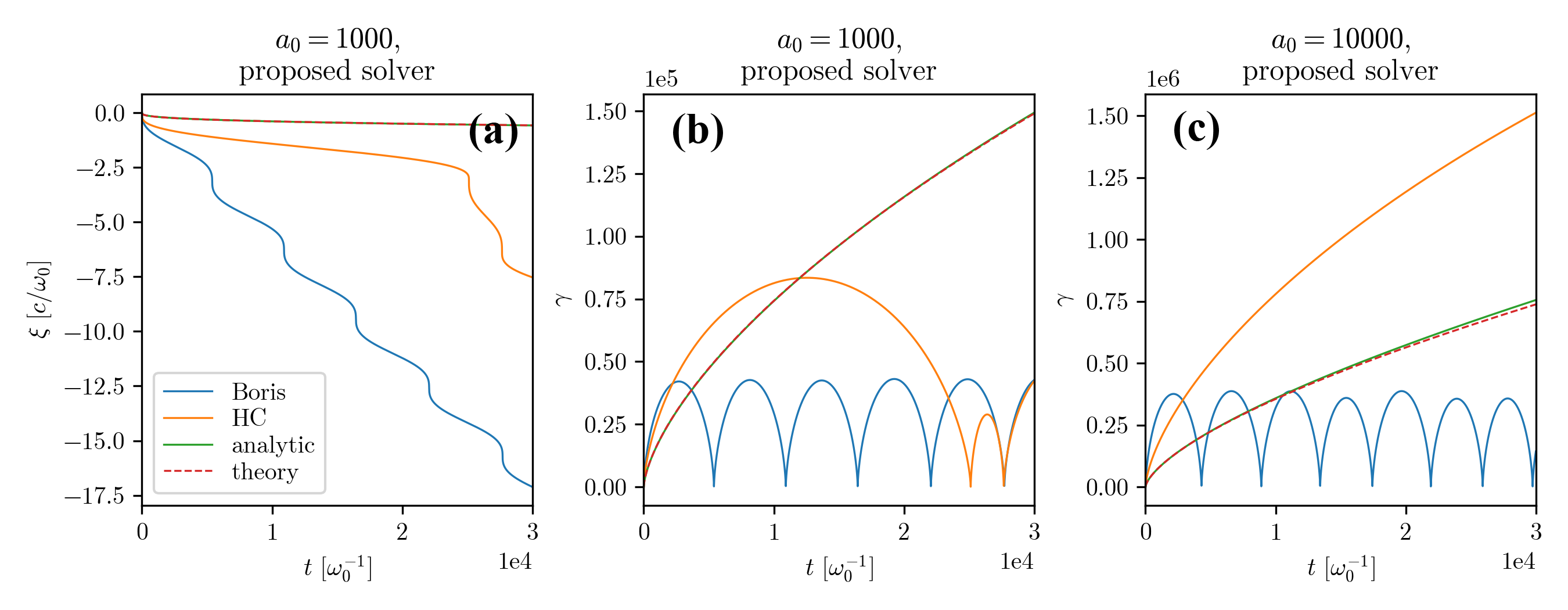}
\caption{Comparison of different pushers combined with the proposed solver for a particle initialized at rest. The evolution of the (a)~particle phase, $\xi=x_1-t$, and (b)~Lorentz factor of a particle in the presence of a laser field with $a_0=1000$. (c)~The evolution of the Lorentz factor of a particle in the presence of a laser field with $a_0=10000$.}
\label{fig:sp3}
\end{figure}

The situation becomes more complicated, however, for larger $a_0$, in which case the numerical errors originating from the particle pusher can be non-negligible. For the test case with $a_0=10$ and using the HC pusher [see Fig.~\ref{fig:sp2}(a)], the phase $\xi$ given by the Yee-type solvers significantly deviates from the theoretical result. The Xu solver gives much better results, and the proposed solver agrees almost perfectly with theory. However, when we increase the laser amplitude to $a_0=100$ [see Fig.~\ref{fig:sp2}(b)], none of the tested solvers give a quantitatively correct result. Nonetheless, the proposed solver still behaves the best among all the tested solvers. The remaining errors are related to the particle pusher. To illustrate this, three pushers (Boris, HC and an analytic pusher that is an extension of the ideas in Gordon et al.~\cite{gordon2017} and  P\'etri~\cite{petri2019}) are tested along with the proposed solver for the $a_0=100$ case. By ``analytic'' pusher, it is meant that an analytic solution is used for the evolution of the proper velocity under the assumption that $\bm E$ and $\bm B$ fields are constant during an interval of time.  It can be seen in Figs.~\ref{fig:sp2}(c) and (d) that only the analytic pusher gives quantitatively correct results. The update to the particle position is not done analytically but is done with second order accurately. However, if the particles are moving near the speed  particle positions these only leads to very small errors. 

\begin{figure}[htbp]
\centering
\includegraphics[width=0.9\textwidth]{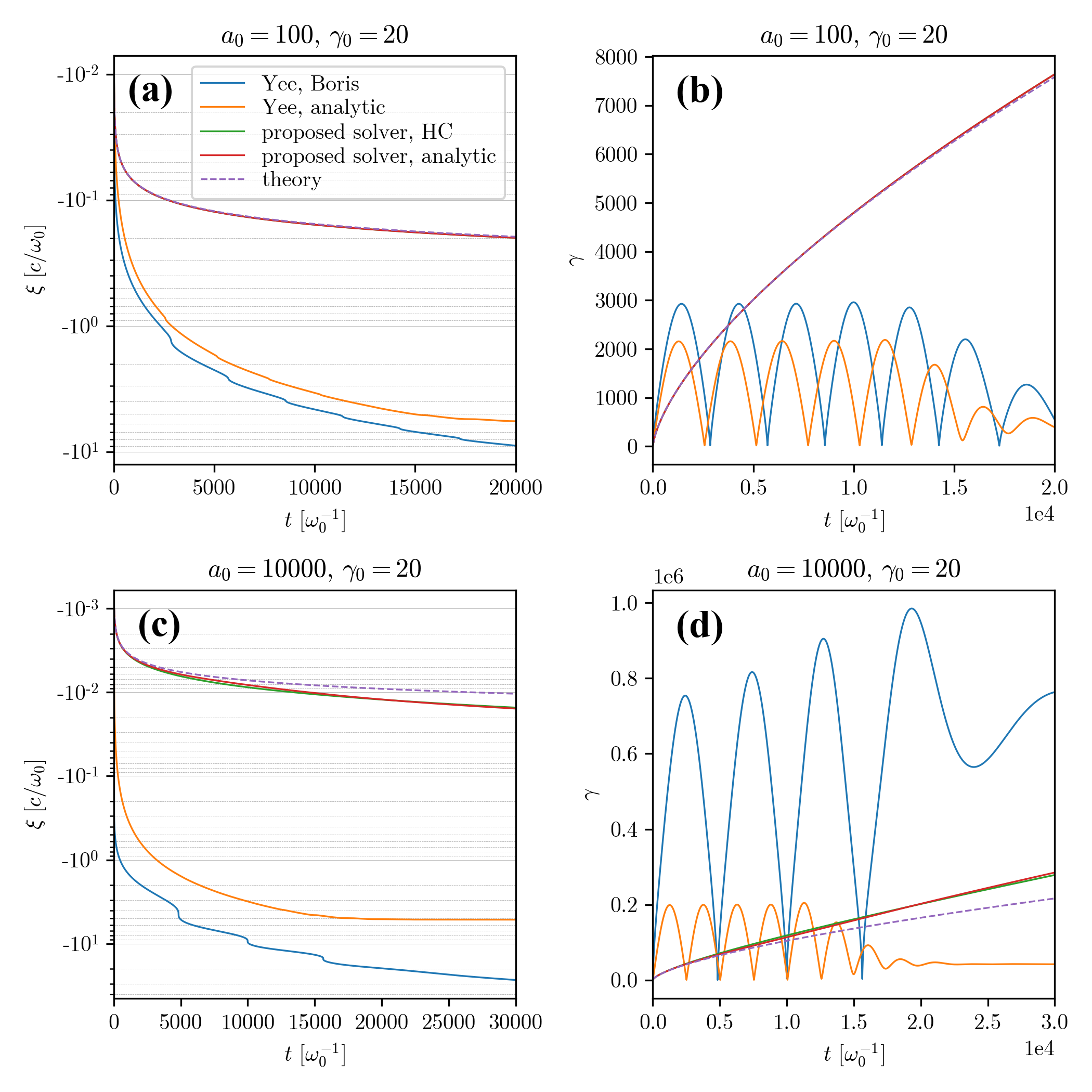}
\caption{Comparison of single-particle trajectories for different solver--pusher combinations for a particle initialized with a drift of $\gamma_0=20$. Evolution of the (a), (c)~particle phase and (b), (d)~Lorentz factor for laser fields of amplitude $a_0=100$ and $a_0=10000$, respectively.}
\label{fig:sp4}
\end{figure}

For even larger laser amplitudes of $a_0=1000$ and $a_0=10000$ (see Fig.~\ref{fig:sp3}), the combination of the proposed solver and analytic pusher still agree well with the theoretical results for a particle initialized at rest, though the use of the Boris and HC pushers introduces significant errors. In Fig.~\ref{fig:sp4}, we explored different solver-pusher combinations for an initially drifting particle with $\gamma_0=20$. For the $a_0=100$ case, the use of the proposed solver combined with both the HC and the analytic pusher achieves excellent agreement with the theory. For the $a_0=10000$ case, these two combinations still work equally well, but significant errors have appeared, indicating that the time step for the pusher and/or the cell size and time step for the field solver may not be small enough. In both cases, note the extreme errors when using the standard Yee solver, even with the analytic pusher.

It should be emphasized that this result may be of great significance for modeling ultra-intense laser and particle interaction. With the onset of petawatt laser systems around the world and multi-petawatt laser systems to be deployed in the near future, experiments are being conducted to examine the complex physics that arises from the interaction between particles and ultra-intense laser fields. The proposed solver and the analytic pusher provide the possibility for high-fidelity simulations in this physics regime for finite-difference-based solvers. We note that the PSATD method should also provide corrections to the numerical dispersion and time-staggering errors. Such a corrected PSATD method combined with the analytic pushers could also provide high-fidelity simulations of particle motion in laser fields. 

\subsection{Synergistic laser wakefield and direct laser acceleration}
\label{sect:simulation:lwfa_dla}

In this section, we will apply the proposed solver to a self-consistent scenario that involves ionized self-injection \cite{pak2010} and acceleration from a combination of wakefields (LWFA) and direct laser acceleration (DLA)~\cite{zhang2015,shaw2017}. We use 2D simulations to illustrate the numerical issues. An ultrafast, intense laser pulse propagates through a neutral gas composed of helium and nitrogen. The helium electrons and the outer-shell electrons of nitrogen are stripped out by the leading front of the laser pulse and form the plasma wake. The electrons in the inner shell of nitrogen are not ionized until they reach the peak intensity of the laser. These inner shell electrons are eventually trapped by the wake and get accelerated. Here, the laser pulse duration is appropriately chosen so that the laser fills the entire first bucket and thus overlaps with the trapped electron bunch. Therefore, the trapped particles are not only accelerated by the longitudinal electric field of the plasma wake, but also may have extra energy gain via a process now known as direct laser acceleration (DLA)~\cite{zhang2015,shaw2017}.

\begin{figure}[htbp]
\centering
\includegraphics[width=0.8\textwidth]{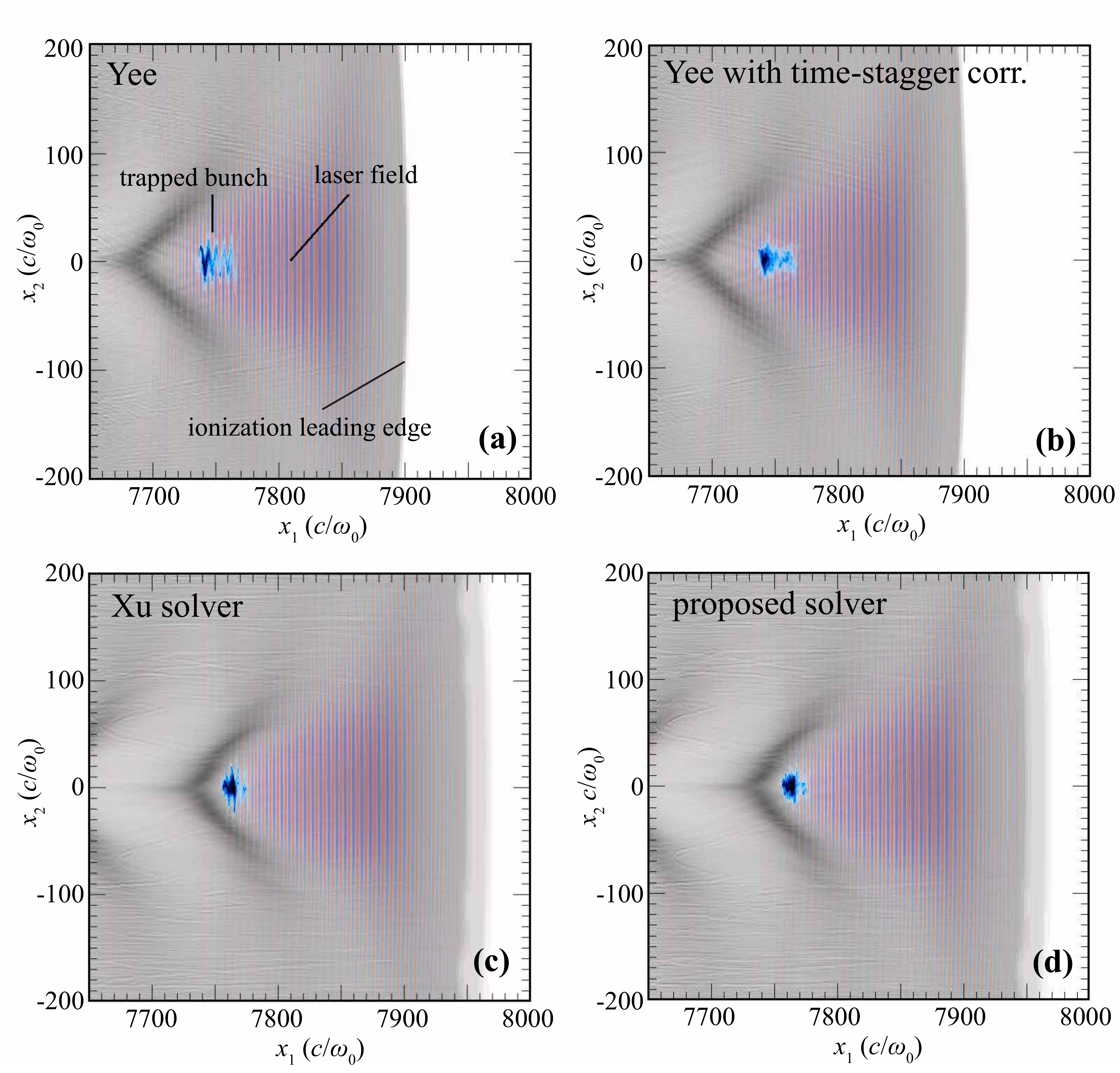}
\caption{2D PIC simulation snapshots when using the (a) standard Yee solver, (b) Yee solver with time-stagger correction, (c) Xu solver and (d) proposed solver. The background electron density distribution, including the helium electrons and the outer-shell nitrogen electrons, is colored gray. The trapped electron bunches are colored blue. The laser electric field is shown in red and blue. The Boris pusher was used for all cases.}
\label{fig:lwfa_dla}
\end{figure}

The simulation parameters are listed in Table~\ref{tab:para_lwfa_dla}. Figure~\ref{fig:lwfa_dla} shows snapshots for each solver around $8000\ k_0^{-1}$ after the laser enters the plasma. Note that the Boris pusher was used for all cases. For Yee-type solvers [see Figs.~\ref{fig:lwfa_dla}(a) and (b)],  the laser pulse travels slower than for those with the Xu and proposed solvers [see Figs.~\ref{fig:lwfa_dla}(c) and (d)] due to their relatively large errors in the dispersion relation (affecting both phase and group velocities). This is evident by comparing the position of witness beams with respect to the wake or of the ionization leading edge. Alternatively, the witness beams in the simulations using solvers without time-stagger correction in the pusher [see Figs.~\ref{fig:lwfa_dla}(a) and (c)] exhibit large spurious modulation in the density distribution. By comparison, for the solvers applying the time-stagger correction [see Figs.~\ref{fig:lwfa_dla}(b) and (d)], the density modulation is greatly mitigated.

\begin{table}[hbtp]
\centering
\begin{tabular}{lll}
\hline\hline
& \textbf{Parameters} & \textbf{Values} \\
\hline
\multirow{4}{*}{Laser} & $a_0$ & 2.0 \\
& wavelength $\lambda_0$ & 0.8 $\mu$m \\
& focal waist $w_0$ & 7 $\mu$m \\
& pulse duration $\tau$ & 45 fs \\
\hline
\multirow{2}{*}{Plasma} & helium density $n_\text{He}$ & $5.0\times10^{18}$ cm$^{-3}$ \\
& nitrogen density $n_\text{N}$ & $1.7\times10^{15}$ cm$^{-3}$ \\
\hline
\multirow{5}{*}{Numerical} & dimension & $(1400,600)k_0^{-1}$ \\
& cell sizes $(\Delta x_1,\Delta x_2)$ & $(0.25,1.0)k_0^{-1}$ \\
& time step $\Delta t$ & $0.125\omega_0^{-1}$ \\
& particles per cell & 8 \\
& particle shape & quadratic \\
\hline\hline
\end{tabular}
\caption{Parameters for synergistic LWFA-DLA simulations.}
\label{tab:para_lwfa_dla}
\end{table}

\begin{figure}[htbp]
\centering
\includegraphics[width=0.6\textwidth]{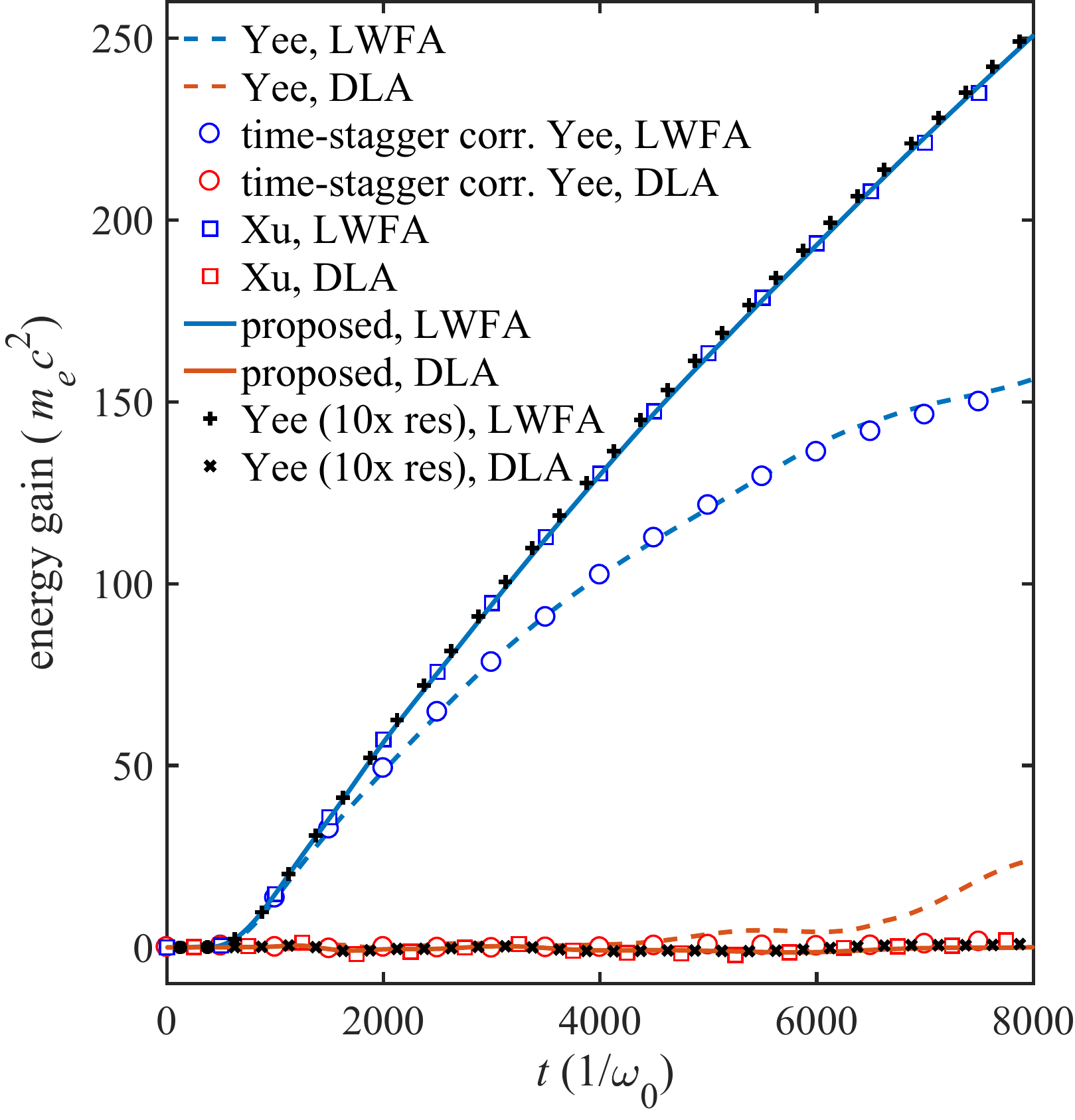}
\caption{Acceleration contribution from the LWFA and DLA mechanisms for different solvers. Note that the LWFA gain for the proposed solver is nearly identical to that of the Yee with 10$\times$ the resolution.}
\label{fig:energy_gain}
\end{figure}

Figure~\ref{fig:energy_gain} shows the acceleration contributed by the LWFA and DLA mechanisms. To generate the energy gain plots, the LWFA and DLA contributions are evaluated by integrating
\begin{equation}
W_\text{LWFA}=-\int_0^t v_1 E_1 dt'
\end{equation}
and
\begin{equation}
W_\text{DLA}=-\int_0^t \bm{v}_\perp \cdot \bm{E}_\perp dt'
\end{equation}
over all time steps. The integrals are averaged over 500 randomly sampled particles from the trapped bunches. Due to the slower phase velocity of the laser resulting from the use of the Yee-type solvers, these trapped bunches undergo faster dephasing in the wake and hence experience a smaller acceleration gradient overall. Specifically for the Yee solver, since the transverse momenta of the trapped bunch are significantly modulated by the spurious force, the DLA contribution is ultimately non-negligible.

The accuracy of the proposed solver was verified by numerical convergence: simulating with the Yee solver using 10 times higher resolution gave LWFA and DLA contributions which converged to those of the proposed solver for larger time steps (and cell sizes). We also investigated the phase space distributions of the accelerated beam and found that only the proposed solver gave out the results converging well to the Yee with 10x higher resolution, while others have larger phase space volume. Therefore, from another perspective, the convergence test indicates that using the proposed solver can give convincing results at a much lower computational cost. Another point learned from the convergence test is, for the selected parameters, the LWFA mechanism dominates the whole acceleration, therefore the numerical dispersion error is important whereas the time-stagger error is a relatively minor issue here.

\section{Conclusion}
\label{sect:conclusion}

In this article, we presented and analyzed three important origins of numerical errors that prevent high-fidelity modeling of the interaction between relativistic charged particles and a co-propagating laser field without the use of small cell sizes and time steps. For a standard FDTD electromagnetic PIC code, errors in (1)~the numerical dispersion relation caused by discretization in time and space, (2)~the Lorentz force induced by advancing the electric and magnetic fields in a time-staggered leap frog algorithm and (3)~the momentum advance in the particle pusher will often lead to significant inaccuracies in the field and particle evolution. To suppress errors from the first two sources, we proposed a novel higher-order finite-difference solver with customized stencil coefficients, which was straightforward to implement into the present framework of the code \textsc{Osiris}. In addition, we compared results with the new solver using the standard Boris, Higuera-Cary and an analytic pusher to demonstrate that the correct choice of the particle pusher can mitigate errors in the momentum advance.

In the proposed Maxwell solver, by introducing different $[k]_1$ operators, \emph{i.e.}, $[k]_{E1}$ into Faraday's law and $[k]_{B1}$ into Ampere's law, the electric and magnetic force felt by a particle in a light wave can be perfectly compensated. With $[k]_{E1}[k]_{B1}=[k]_{1,t}^2$, which yields the true dispersion relation $\omega=ck_1$, the proposed solver is nearly free of numerical dispersion errors for a laser propagating in the $\hat{1}$-direction. Since the charge-conserving current deposition scheme is only suitable for the second-order-accurate solver in the present \textsc{Osiris}, we modified the current deposition appropriately for the proposed solver with wider stencils. We have shown that by correcting the current in Fourier space, both the continuity equation and Gauss's law remain satisfied at each time step.

The advantages of the proposed solver have been verified by two sets of simulations: (1)~a single particle co-propagating with a plane wave laser and (2)~LWFA with ionization injection and DLA. It is shown that the proposed solver can yield results close to analytic solutions, while standard solvers can distort the physics or even lead to  incorrect results. The choice of the particle pusher was also shown to be important and can lead to additional errors. The use of the proposed solver---in conjunction with an accurate particle pusher---enables high-fidelity simulations of particle motion in ultra-intense laser fields. The analysis described is also useful for standard FFT and PSATD algorithms.

\section*{Acknowledgments}
This work was supported in parts by the  US Department of Energy contract number  DE-SC0010064 and  SciDAC FNAL subcontract 644405, Lawrence Livermore National Laboratory subcontract B634451 and US National Science Foundation grant number 1806046. Simulations were carried out on the Cori Cluster of the National Energy Research Scientific Computing Center (NERSC).

\begin{appendix}
\section{Lorentz force exerted on a macro-particle}
\label{sect:app:force}

Without loss of generality, we consider the two-dimensional case with a macro-particle of charge $q$ described by the continuous coordinates $(x_1,x_2)$. The transverse component of Lorentz force felt by the particle is interpolated from $E_2$ and $B_3$ defined on the discrete grid points as
\begin{align}
\begin{split}
\frac{F_2(t^n,x_1,x_2)}{q}&=\sum_{i_1,i_2}E_{2,i_1,i_2+\frac{1}{2}}^n S\left(X_1^{i_1}-x_1,X_2^{i_2+\frac{1}{2}}-x_2\right) \\
&-\bar{\beta}_1\frac{B_{3,i_1+\frac{1}{2},i_2+\frac{1}{2}}^{n-\frac{1}{2}}+B_{3,i_1+\frac{1}{2},i_2+\frac{1}{2}}^{n+\frac{1}{2}}}{2} S\left(X_1^{i_1+\frac{1}{2}}-x_1,X_2^{i_2+\frac{1}{2}}-x_2\right),
\end{split}
\end{align}
where $t^n=n\Delta t$ is discrete time, $X_k^{i_k}=i_k\Delta x_k$ is the position of spatial grid points in the $\hat{k}$-direction and $S$ is the interpolation function. Note that we have already considered the spatial staggering of $E_2$ and $B_3$.

After a Fourier transform we have
\begin{align}
\begin{split}
\frac{\tilde{F}_2}{q}&=\sum_{i_1,i_2} E_{2,i_1,i_2+\frac{1}{2}}^n \tilde{S}(-k_1,-k_2)\exp\left(j\omega t^n-jk_1 X_1^{i_1}-jk_2 X_2^{i_2+\frac{1}{2}}\right) \\
&-\bar{\beta}_1 \frac{B_{3,i_1+\frac{1}{2},i_2+\frac{1}{2}}^{n-\frac{1}{2}}+B_{3,i_1+\frac{1}{2},i_2+\frac{1}{2}}^{n+\frac{1}{2}}}{2} \tilde{S}(-k_1,-k_2)\exp\left(j\omega t^n-jk_1 X_1^{i_1+\frac{1}{2}}-jk_2 X_2^{i_2+\frac{1}{2}}\right) \\
&=\left[\tilde{E}_2(\omega,k_1,k_2)-\bar{\beta}_1 \tilde{B}_3(\omega,k_1,k_2)\cos\frac{\omega \Delta t}{2}\right] \tilde{S}(-k_1,-k_2).
\end{split}
\end{align}

\section{Approximate $[k]_{E1}$ and $[k]_{B1}$ using stencil coefficient customization}
\label{sect:app:solver}

In this appendix, we follow the method in Ref.~\cite{li2017} to construct discrete operators $\tilde{[k]}_{E1}$ and $\tilde{[k]}_{B1}$ that best approximate the desired $[k]_{E1}$ and $[k]_{B1}$ operators. Their corresponding solver is assumed to have $p^\text{th}$-order accuracy for the partial derivative in the $\hat{1}$-direction. In Faraday's law, the finite-difference operator for the partial derivative in $x_1$ can be written as
\begin{equation}
\partial^E_{x_1}f_{i_1,i_2}=\frac{1}{\Delta x_1}\sum_{j=1}^M C_j^E(f_{i_1+j,i_2}-f_{i_1-j+1,i_2}).
\end{equation}
Similarly, for Ampere's law, we have
\begin{equation}
\partial^B_{x_1}f_{i_1,i_2}=\frac{1}{\Delta x_1}\sum_{j=1}^M C_j^B(f_{i_1+j-1,i_2}-f_{i_1-j,i_2}).
\end{equation}
Performing a Fourier transform, the corresponding operators in $k$-space become
\begin{equation}
\tilde{[k]}_{E1,B1}=\sum_{j=1}^M C_j^{E,B}\frac{\sin[(2j-1)k_1\Delta x_1/2]}{\Delta x_1/2}.
\end{equation}
For a standard high-order operator, the number of coefficients $M=p/2$. But here we need to extend the stencil ($M>p/2$) to obtain more degrees of freedom for the purpose of fitting the given $[k]_{E1}$ and $[k]_{B1}$. To simplify the notations, we normalize $[k]_{E1,B1}$, $\tilde{[k]}_{E1,B1}$ and $k_1$ to $k_{g1}=2\pi/\Delta x_1$ herefrom. In the spirit of the least squares approximation, a function such as
\begin{equation}
\mathcal{F}=\int_0^{1/2}w(k_1)(\tilde{[k]}_{E1,B1}-[k]_{E1,B1})^2 dk_1
\end{equation}
should be minimized to obtain the stencil coefficients, where $w(k_1)$ is the weight function and $\tilde{[k]}_{E1,B1}$ is the approximation. In addition, the discrete operator is subject to the constraint $\partial^{E,B}_{x_1}\rightarrow \partial_{x_1}+O(\Delta x_1^p)$, which can be guaranteed by the matrix equation $\bm{\mathcal{M}}\bm{C}^{E,B}=\bm{e}_1$, where $\bm{C}^{E,B}\equiv(C_1^{E,B},\ldots,C_M^{E,B})^T$, $\bm{e}_1\equiv(1,0,\ldots,0)^T$ and the matrix element $\mathcal{M}_{ij}=(2j-1)^{2i-1}/(2i-1)!$ with $i=1,\ldots,p/2$ and $j=1,\ldots,M$. Specifically for the second-order accuracy used throughout this paper, $\bm{\mathcal{M}}$ reduces to a row vector with elements $\mathcal{M}_j=2j-1$.

We introduce the Lagrangian
\begin{equation}
\mathcal{L}=\mathcal{F}+\bm{\lambda}^T(\bm{\mathcal{M}}\bm{C}^{E,B}-\bm{e}_1)
\end{equation}
to solve the constrained least-squares minimization problem, where $\bm{\lambda}=(\lambda_1,\ldots,\lambda_{p/2})^T$ is a Lagrange multiplier. The stencil coefficients can be found out by seeking extrema of $\mathcal{L}$, \emph{i.e.},
\begin{equation}
\frac{\partial\mathcal{L}}{\partial C_j^{E,B}}=0,\ j=1,\ldots,M\quad
\text{and}\quad
\frac{\partial\mathcal{L}}{\partial \lambda_i}=0,\ i=1,\ldots,p/2.
\end{equation}
This can be written into a matrix equation
\begin{equation}
\begin{pmatrix}
\bm{\mathcal{A}} & \bm{\mathcal{M}}^T \\
\bm{\mathcal{M}} & \bm{0}
\end{pmatrix}
\begin{pmatrix}
\bm{C}^{E,B} \\
\bm{\lambda}
\end{pmatrix}=\begin{pmatrix}
\bm{b}^{E,B} \\ \bm{e}_1
\end{pmatrix},
\end{equation}
where $\bm{\mathcal{A}}$ is an $M\times M$ matrix and $\bm{b}^{E,B}$ is an $M$-dimensional column vector, each with elements
\begin{align}
\mathcal{A}_{ij}&=\frac{2}{\pi^2}\int_0^{1/2}w(k_1)\sin[(2i-1)\pi k_1]\sin[(2j-1)\pi k_1] dk_1, \\
b_i^{E,B}&=\frac{2}{\pi}\int_0^{1/2}w(k_1)\sin[(2i-1)\pi k_1][k]_{E1,B1}dk_1.
\end{align}
Mathematically, it is usually impossible to approximate the target operators uniformly well in the whole primary Brillouin zone, $k_1\in[0,1/2]$. Therefore, a proper weight function $w(k_1)$ is needed for relaxation. To ensure accurate fit in the low- and moderate-$k_1$ regions with only loose requirement in the high-$k_1$ region, we can use a super-Gaussian weight function
\begin{equation}
w(k_1)=\exp\left[-\ln(2)\left(\frac{2k_1}{w_{k1}}\right)^n\right],
\end{equation}
where $n$ is an integer and $w_{k1}$ specifies the super-Gaussian width. In practice, we often use $n=10$ and $w_{k1}=0.3$--0.4.

\section{Numerical dispersion relation in cold plasma}
\label{sect:app:nd}

We follow the theoretical framework and notation in reference~\cite{xu2013} to derive the numerical dispersion relation. From Eq.~(\ref{eq:maxwell_fourier}) we can obtain
\begin{align}
\begin{split}
\label{eq:plasma_nd}
&\left([\omega]_t^2-[\bm{k}]_E\cdot[\bm{k}]_B+[\bm{k}]_E[\bm{k}]_B\right)\tilde{\bm{E}}=-i\tilde{\bm{J}} \\
=&-\omega_p^2\sum_{\mu,\bm{\nu}}(-1)^\mu\left\{\int\frac{\tilde{\bm{S}}_J(-\bm{k}')}{\gamma\omega'-\bm{k}'\cdot\bm{p}}\bm{p}\left[[\omega]_t\tilde{\bm{S}}_E(\omega',\bm{k}')\tilde{\bm{E}} \right.\right. \\
&\left.\left.+\frac{\bm{p}}{\gamma}\times\tilde{\bm{S}}_B(\omega',\bm{k}')([\bm{k}]_E\times\tilde{\bm{E}})\right]\cdot\PP{f_0}{\bm{p}}d^3\bm{p}\right\}
\end{split}
\end{align}
where $\omega_p$ is the plasma frequency, $\tilde{\bm{S}}_Q$ is the Fourier-transformed interpolation tensor for field $\bm{Q}$, $f_0$ is the equilibrium distribution function for the plasma and $(\omega',\bm{k}')$ is defined as
\begin{align}
\omega'&\equiv\omega+\mu\omega_g,\quad\omega_g=2\pi/\Delta t, \quad\mu=0,\pm1,\pm2,\dots \nonumber \\
k'_i&\equiv k_i+\nu_i k_{gi}, \quad k_{gi}=2\pi/\Delta x_i, \quad\nu_i=0,\pm1,\pm2,\dots \nonumber
\end{align}
The expression for $\tilde{\bm{J}}$ is given by use of the linearized Vlasov equation after Fourier transform (See reference~\cite{xu2013} for the derivation). We can finally rewrite Eq.~(\ref{eq:plasma_nd}) into the matrix form
\begin{equation}
\nonumber
\bm{\epsilon}(\omega,\bm{k})\bm{E}=\bm{0},
\end{equation}
and the numerical dispersion relation can be found by vanishing the determinant of $\bm{\epsilon}$, which is similar to the dielectric tensor.

We are interested in a uniform, cold plasma with equilibrium distribution function $f_0=\delta(p_1)\delta(p_2)\delta(p_3)$. Note that $f_0$ is normalized to the plasma density as is the original definition in reference~\cite{xu2013}. If we substitute $f_0$ into Eq.~(\ref{eq:plasma_nd}) and conduct the integration, we can obtain all the elements in the tensor $\bm{\epsilon}$. Instead of giving out the tedious full set of matrix elements for the 3D case, we can learn much from the 2D limit without loss of generality. It can be shown that the elements of $\bm{\epsilon}$ in the 2D limit are
\begin{align}
\epsilon_{11} &= [\omega]_t^2-[k]_{E2}[k]_{B2}-\omega_p^2\sum_{\mu,\bm{\nu}}(-1)^\mu \tilde{S}_{J1}\tilde{S}_{E1}\frac{[\omega]_t}{\omega'} \nonumber \\
\epsilon_{12} &= [k]_{E1}[k]_{B2} \nonumber \\
\epsilon_{21} &= [k]_{E2}[k]_{B1} \nonumber \\
\epsilon_{22} &= [\omega]_t^2-[k]_{E1}[k]_{B1}-\omega_p^2\sum_{\mu,\bm{\nu}}(-1)^\mu \tilde{S}_{J2}\tilde{S}_{E2}\frac{[\omega]_t}{\omega'}  \nonumber \\
\epsilon_{33} &= [\omega]_t^2-[k]_{E1}[k]_{B1}-[k]_{E2}[k]_{B2}-\omega_p^2\sum_{\mu,\bm{\nu}}(-1)^\mu \tilde{S}_{J3}\tilde{S}_{E3}\frac{[\omega]_t}{\omega'}, \nonumber
\end{align}
and all other elements vanish. According to the condition $\text{det}(\bm{\epsilon})=0$, the numerical dispersion relation is determined by $\epsilon_{11}\epsilon_{22}-\epsilon_{12}\epsilon_{21}=0$ and $\epsilon_{33}=0$. Viewing these two equations in the continuous limit, the former actually gives the dispersion relation as the product of both Langmuir and electromagnetic modes, \emph{i.e.}, $(\omega^2-\omega_p^2)(\omega^2-k^2-\omega_p^2)=0$; note that in the discrete scenario the two modes are generally coupled together. The latter gives the dispersion relation for the electromagnetic mode. If we define $s_i\equiv\sum_{\mu,\bm{\nu}}(-1)^\mu\tilde{S}_{Ji}\tilde{S}_{Ei}\frac{[\omega]_t}{\omega'}$ to simplify the notation, the equation $\epsilon_{11}\epsilon_{22}-\epsilon_{12}\epsilon_{21}=0$ can be written as
\begin{equation}
\begin{split}
\label{eq:plasma_nd2}
\left([\omega]_t^2-\frac{[\bm{k}]_E\cdot[\bm{k}]_B+(s_1+s_2)\omega_p^2+\sqrt{\Delta}}{2}\right) \times \\
\left([\omega]_t^2-\frac{[\bm{k}]_E\cdot[\bm{k}]_B+(s_1+s_2)\omega_p^2-\sqrt{\Delta}}{2}\right) = 0,
\end{split}
\end{equation}
where
\begin{equation}
\nonumber
\Delta=([\bm{k}]_E\cdot[\bm{k}]_B)^2-2\omega_p^2(s_1-s_2)([k]_{E1}[k]_{B1}-[k]_{E2}[k]_{B2})+(s_1-s_2)^2\omega_p^4.
\end{equation}
It can be easily verified that, for the continuous limit ($s_i\rightarrow1$), the first term in Eq.~(\ref{eq:plasma_nd2}) reduces to the dispersion relation of the electromagnetic mode and the second term corresponds to that of the Langmuir mode. This numerical dispersion relation can be further simplified if we assume $S_{Ji}=S_{Ei}$. By referring to the explicit expression of the interpolation tensor in the appendix of Ref.~\cite{xu2013}, we have $s_1=s_2\triangleq s$ for a momentum-conserving scheme. In this case, $\Delta=[\bm{k}]_E\cdot[\bm{k}]_B$ and thus Eq.~(\ref{eq:plasma_nd2}) becomes
\begin{equation}
\label{eq:plasma_nd_simple}
\left([\omega]_t^2-[\bm{k}]_E\cdot[\bm{k}]_B+s\omega_p^2\right)
\left([\omega]_t^2-s\omega_p^2\right) = 0.
\end{equation}
Now the numerical dispersion relation given by $\epsilon_{11}\epsilon_{22}-\epsilon_{12}\epsilon_{21}=0$ is identical to that given by $\epsilon_{33}=0$.

\end{appendix}

\bibliographystyle{elsarticle-num}
\bibliography{refs}

\end{document}